\documentclass[3p]{elsarticle}

\usepackage{amsmath}
\usepackage{graphicx}
\usepackage{subcaption}
\usepackage{epstopdf}
\usepackage{mathtools}
\usepackage{upgreek}
\usepackage{placeins}

\DeclareMathOperator{\sgn}{sgn}

\journal{ }

\biboptions{authoryear}

\begin{document}

\begin{frontmatter}

\title{A mechanistic model for the growth of cylindrical debris particles in the presence of adhesion}

\author[epfl]{Enrico Milanese}
\author[epfl]{Jean-Fran\c{c}ois Molinari\corref{mycorrespondingauthor}}
\cortext[mycorrespondingauthor]{Corresponding author}
\ead{jean-francois.molinari@epfl.ch}

\address[epfl]{Civil Engineering Institute, Materials Science and Engineering Institute, \'{E}cole Polytechnique F\'{e}d\'{e}rale de Lausanne (EPFL), CH-1015 Lausanne, Switzerland}




\begin{abstract}
The wear volume is known to keep increasing during frictional processes, and Archard notably proposed a model to describe the probability of wear particle formation upon asperity collision in a two-body contact configuration. While this model is largely adopted in the investigations of wear, the presence of wear debris trapped between the surfaces changes the system into a three-body contact configuration already since the early stages of the process. In such a configuration, a significant amount of wear is produced at the interface between the trapped debris and the sliding bodies. Here, relying on analytical models, we develop a framework that describes crack growth in a three-body configuration at the particle-surface interface. We then show that crack growth is favoured within the sliding surfaces, instead of within the debris particle, and test such result by means of numerical simulations with a phase-field approach to fracture. This leads to an increase in the wear volume and to debris particle accretion, rather than its break down. The effects of adhesion, coefficient of friction, and ratio of the applied global tangential and normal forces are also investigated.
\end{abstract}


\end{frontmatter}

\section{Introduction}

The importance of wear for the performance and durability of mechanical components has been known for centuries, as the first documented studies of the topic date back to~\cite{davincicodexI}. While his contributions saw the light only recently and are still to be clearly assessed~\citep{hutchings2016leonardo}, he observed that the amount of wear was larger when the sliding motion was longer~\citep{davincicodexI}, and that harder materials would wear less~\citep{davinci1519atlanticus}. The link between these quantities was made much clearer by~\cite{holm1946} and~\cite{archard1953contact} in the past century, at least for the case of adhesive wear. Archard formulated the wear law that still carries his name and that states that the wear volume is linearly proportional to the sliding distance, the normal load, and the inverse of the hardness of the materials. The proportionality is governed by the wear coefficient, which is generally constant and is determined experimentally. According to Archard's picture, the wear volume steadily increases during adhesive wear processes, and the wear coefficient describes the probability of two asperities belonging to opposite rough surfaces to form a debris particle upon collision.

In recent years, extensive experimental~\citep{merkle2008liquid,liu2010preventing,liu2010method,bhaskaran2010ultralow,jacobs2013nanoscale,yang2016adhesion} and numerical~\citep{sorensen1996simulations,aghababaei2016critical,aghababaei2017debris,aghababaei2018asperity} investigations have been carried out to shed light on what happens during such asperity contacts or collisions, and three possible mechanisms have been identified. When both the normal load and the adhesion forces are low, atom-by-atom removal takes place~\citep{bhaskaran2010ultralow,jacobs2013nanoscale,yang2016adhesion}, while for larger loads and larger adhesion forces, asperities mutually deform plastically~\citep{merkle2008liquid,aghababaei2016critical}, or break in a brittle fashion~\citep{liu2010preventing,liu2010method,aghababaei2016critical}. The transition from the ductile to the brittle behaviour is determined by a material-dependent critical length scale~\citep{aghababaei2016critical}. In the brittle scenario, the asperities form a debris particle with a well-defined initial volume, which is proportional to the tangential load~\citep{aghababaei2017debris}. It has been hypothesized that Archard's wear coefficient includes the probability of each asperity junction being smaller or larger than the aforementioned critical length scale, and efforts are made towards a mechanical description of the wear coefficient~\citep{Frerot2018}.

This body of work focuses though on two-body contacts, that is two surfaces (the first bodies) come directly into contact at the asperity level. As soon as a loose debris particle forms, this constitutes the third body in the system and it is trapped between the two first bodies. Three-body contact then takes place locally: the debris particle separates the two surfaces, which are each in contact with the particle, and not directly with one another. Such transition to a three-body contact has been recently found to be key, for instance, in the evolution of the sliding surfaces into rough self-affine morphology~\citep{milanese2019emergence}. It is arguable that loose debris particles form since the early stages, and that in the study of wear a third-body approach is more suitable~\citep{godet1984third,berthier1988velocity,descartes2002rheology,fillot2007wear}. According to this approach, a load-bearing film of rolling third bodies is responsible for the changes in the rheology of the system, and behaves as a lubricant. This approach is supported by experimental~\citep{cocks1962interaction,harris2015wear,hintikka2017third} and numerical~\citep{fillot2005simulation,fillot2007modelling,renouf2011numerical,milanese2019emergence} evidence: the transition to three-body contact is linked, for instance, to a reduction of the wear rate~\citep{fillot2005simulation,fillot2007modelling,harris2015wear,hintikka2017third,milanese2019emergence}. Similarly, a decrease in gouge-formation rates has been observed in natural faults~\citep{brodsky2011faults} and rock experiments~\citep{boneh2013frictional}. According to some characteristics of the system (e.g.\ surface roughness, system size, wear evacuation rate, wear production rate, debris particle size), the portion of the system where contact happens in a three-body configuration can be more or less spread.

\begin{figure*}
  \centering
  \includegraphics[width=.95\textwidth]{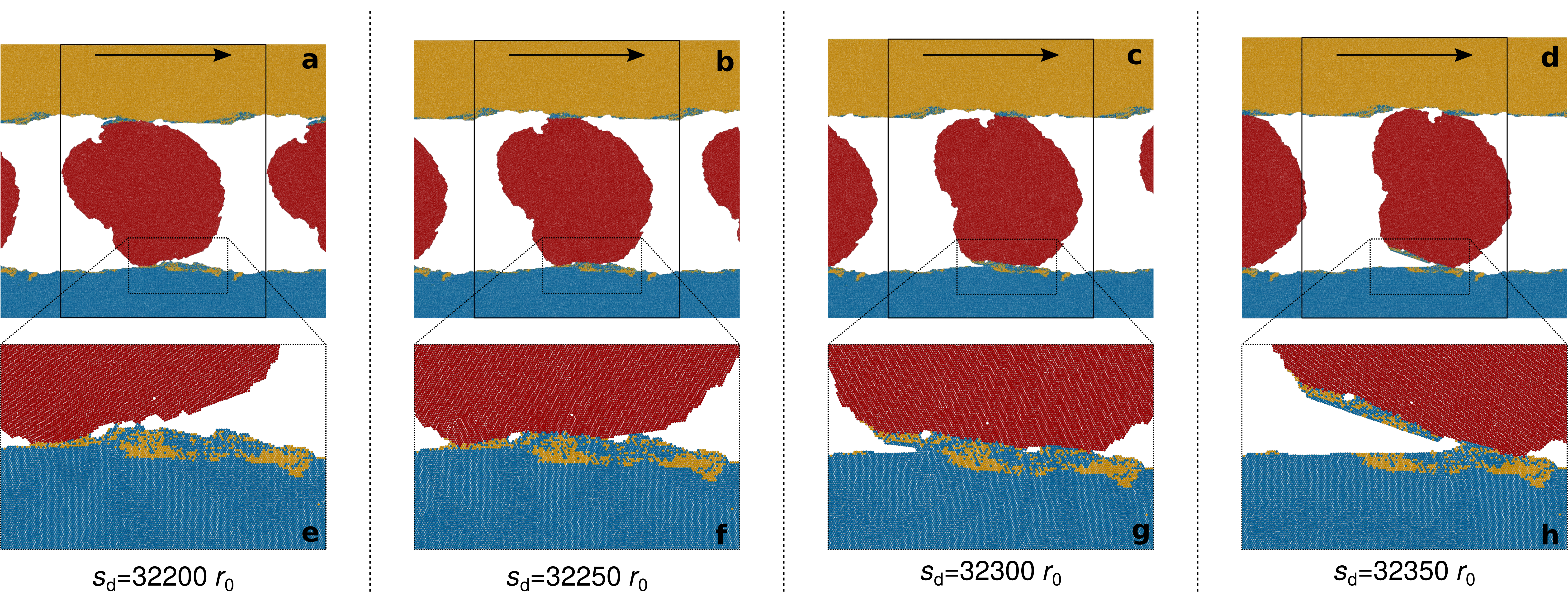}
  \caption{Example of particle accretion observed in 2D molecular dynamics simulations. a-d) The debris particle (identified by red atoms) is rolling between the two first bodies (a) and adheres to the bottom surface (b); upon continuous rolling, the trailing edge is being pulled away (c) and material is transferred from the surface to the debris particle (d). e-h) Magnification of the areas within the dotted black rectangles in panels (a-d). In all panels, colours distinguish atoms originally belonging to the top (yellow) and bottom (blue) bodies. In panels (a-d), solid black lines represent simulation box boundaries (periodic boundary conditions are applied), and the arrow indicates the sliding direction of the top body (the bottom body is fixed). Snapshots are taken at the sliding distance $s_\mathrm{d}$ expressed in multiples of the atoms equilibrium distance $r_0$. Atom interactions are all described by the same pair potential (i.e.\ full adhesion conditions are recovered upon contact). Snapshot are taken from simulation R1 of \cite{milanese2019emergence}, the reader is referred to such work for more details.}
  \label{fig:md}
\end{figure*}

Finally, recent numerical simulations~\citep{milanese2019emergence} have shown that the volume of a debris particle trapped between two surfaces overall increases with time (or, equivalently, the sliding distance), and particle accretion is generally favoured over deposition of fragments from the particle onto the surfaces. Inspired by such observations (see Figure~\ref{fig:md}), we focus here on the wear production during three-body contact, within a solid mechanics framework. In particular, we show that the stress field in a three-body contact configuration favours an increase in the wear volume. To this end, the framework of our approach is presented in Section~\ref{sec:theory}, and in Section~\ref{sec:calculation} the existing contact solutions relevant to our problem are briefly presented, and they are used to determine the stress field in rolling contact with and without adhesion. In Section~\ref{sec:results}, the results of such approach are discussed, and tested with a numerical method. The conclusions are finally reported in Section~\ref{sec:conclusions}.

\begin{table*}
  \centering
  \begin{tabular}{c c p{8.cm} c}
    Symbol & Expression & Description & Physical dimension \\[2pt]
    \hline \\[-8pt]
    $\xi$, $\eta$, $\zeta$ & $x/l$, $y/l$, $z/l$ & non-dimensional $x$, $y$, and $z$ coordinate & --- \\[4.5pt]
    $a$, $b$, $c$  & --- & half-width of contact, adhesion, and interaction & length \\[4.5pt]
    $\alpha$, $\beta$, $\chi$  & $a/l$, $b/l$, $c/l$ & non-dimensional half-width of contact, adhesion, and interaction & --- \\[4.5pt]
    $m$ & $\chi/\alpha$ & scaled non-dimensional interaction half-width & --- \\[4.5pt]
    $\lambda$ & $\sigma_0/s$ & non-dimensional limiting adhesive stresses (Maugis parameter) & --- \\[4.5pt]
    $l$ & $\left( \frac{R^2 w}{E^*} \right)^{1/3}$ & non-dimensionalizing length & length \\[4.5pt]
    $s$ & $\left( \frac{w E^{*2}}{R} \right)^{1/3}$ & non-dimensionalizing pressure & force/area \\[4.5pt]
    $P_0$ & $\left( R w^2 E^* \right)^{1/3}$ & non-dimensionalizing load & force/length \\[4.5pt]
    $R$ & --- & cylindrical particle radius & length \\[4.5pt]
    $w$ & --- & fracture energy & energy/area \\[4.5pt]
    $E^*$ & --- & effective elastic modulus & force/area \\[4.5pt]
    $P$, $T$ & --- & normal and tangential component of the transmitted force per unit length & force/length \\[4.5pt]
    $Z(x)$, $X(x)$ & --- & transmitted normal and tangential tractions & force/area \\[4.5pt]
    $X'(x)$, $X''(x)$ & --- & components of the transmitted tangential tractions & force/area \\[4.5pt]
    $\mu$ & --- & Coulomb coefficient of friction & --- \\[4.5pt]
    $\sigma_{ij}$ & --- & Cauchy stress & force/area \\[4.5pt]
    $\sigma_{ij}^\mathrm{n}$ & --- & Cauchy stresses due to $Z(x)$ & force/area \\[4.5pt]
    $\sigma_{ij}^\mathrm{t}$ & --- & Cauchy stresses due to $X(x)$ & force/area \\[4.5pt]
    $\sigma_{1,2}$ & --- & maximum (1) and minimum (2) principal Cauchy stresses on the $<\xi,\zeta>$ plane & force/area \\[4.5pt]
    $\sigma_0$ & --- & limiting adhesive stresses & force/area \\[4.5pt]
    $\tilde{\sigma}_\mathrm{c}$ & --- & limiting non-dimensional tensile stresses & --- \\[4.5pt]
    $ \tilde{\left( \cdot \right)}$ & --- & non-dimensional loads and pressures & --- \\[4.5pt]
    $ \bar{\left( \cdot \right)}$ & --- & quantities computed at $\zeta = 0$ & --- \\[4.5pt]
    $\theta_\mathrm{p}$ & --- & angle between the $\xi$ axis and the direction of the maximum principal stress & --- \\[4.5pt]
    $\theta_\mathrm{c}$ & $\theta_\mathrm{p} - \uppi/2$ & angle between the $\xi$ axis and the direction of crack propagation & --- \\[4.5pt]
    $\rho$ & --- & non-dimensional half-width over which $X''$ is applied & --- \\[4.5pt]
    $\psi$ & --- & non-dimensional coordinate of the midpoint of $2\rho$ & --- \\[4.5pt]
    $\tilde{p}_0$ & $2 \tilde{P} / \uppi \alpha$ & maximum value of $\tilde{Z}(\xi)$ at the center of the contact width & --- \\[4.5pt]
    $\tilde{X}_\mathrm{adh}$ & --- & tangential tractions due to adhesive friction & --- \\[4.5pt]
    $\tilde{\tau}_0$ & $\tilde{X}_\mathrm{adh}/2\chi$ & non-dimensional adhesive friction tangential stress & --- \\[4.5pt]
    $\tilde{\tau}_\mathrm{c}$ & --- & non-dimensional shear strength of the contact interface & --- \\
    \hline
  \end{tabular}
  \caption{Nomenclature of the main symbols.}
  \label{tab:nomenclature}
\end{table*}

\section{Theory}\label{sec:theory}

\subsection{Approach}\label{ssec:app}

In the simplest three-body contact, one wear debris particle is rolling between two surfaces, while in contact with both of them. The core idea of our approach is to view the rolling motion of the particle as the opening and closure of two cracks. At the leading edge of the contact, the rolling motion is closing the crack, while at the trailing edge the crack opens. This approach is inspired by the work of~\cite{maugis1992adhesion,maugis2000modern}, who used a fracture mechanics approach to investigate the adhesive contact of spheres.

In the remaining of the manuscript we focus on the contact between the particle and one of the surfaces, as the behaviour at the contact with the other surface is quickly derivable with symmetry arguments. Our approach then implies that the direction of propagation of the opening crack at the trailing edge of contact determines which of the bodies wears out: the particle or the surface.

In the following Sections, we derive the stress fields due to the rolling contact with adhesion, the principal stresses and the crack propagation angle at the trailing edge based on the maximum hoop stress criterion~\citep{erdogan1963crack}.

\subsection{Framework}\label{ssec:fw}

\begin{figure*}
  \centering
  \includegraphics[width=\textwidth]{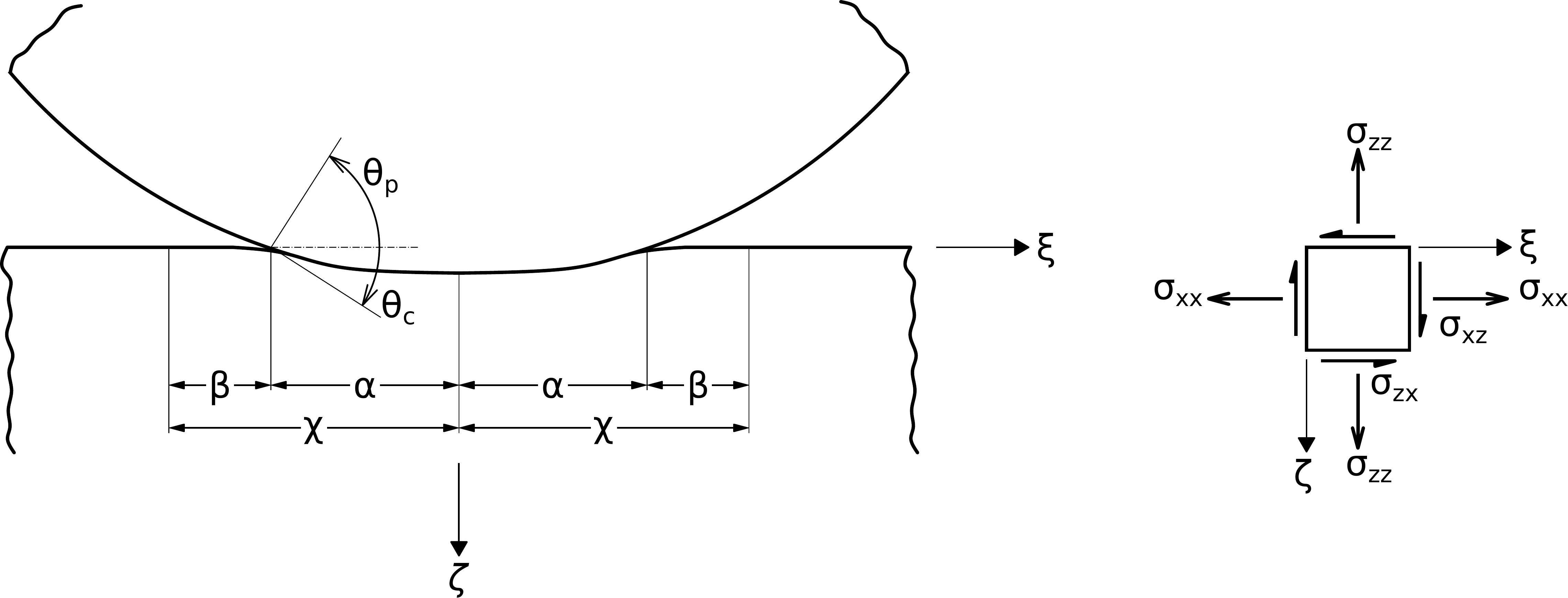}
  \caption{Frame of reference for a body in contact with a half-space (left) and stress directions taken as positive for a small representative square element at a random point within the bodies (right). Length are non-dimensionalized by $l$ (see Equations~\ref{eq:nondim_a} to~\ref{eq:nondim_l} ): $\xi=x/l$, $\zeta=z/l$. Note that the signs of the shear stresses are the opposite than in the traditional continuum solid mechanics frame of reference. Normal stresses are positive when tensile. The $\eta$-axis, with $\eta=y/l$, is not displayed and is orthogonal to the $<\xi,\zeta>$ plane and pointed toward the reader. $\theta_\mathrm{p}$ is the angle between the direction of the largest principal stress and the axis $\xi$, $\theta_\mathrm{c} = \theta_\mathrm{p} - \uppi/2$ is the angle between the axis $\xi$ and the crack path (see Section~\ref{ssec:cangle}). Angles are positive if measured counter-clock wise.}
  \label{fig:fw}
\end{figure*}

We consider the case of two-dimensional contact between two cylinders with their longitudinal axis $y$ aligned (see Figure~\ref{fig:fw}). In our case we assume that one of the two cylinders has infinite radius of curvature and represents the surface against which the debris particle is rolling.

Three different types of rolling motion can take place in such conditions~\citep{johnson1958effect,johnson1987contact}:
\begin{itemize}
\item free rolling: if the transmitted resultant force is normal to the contact width;
\item tractive rolling: if the transmitted resultant force is not normal to the contact width (i.e.\ tangential forces are transmitted, too);
\item rolling with spin: if a relative angular velocity along the axis normal to the contact width ($\zeta$-axis in Figure~\ref{fig:fw}) exists between the cylinder and the surface;
\end{itemize}
we will consider here the first two types of rolling.

\paragraph{Frame of reference and formalism}

Following the common conventions of contact mechanics~\citep{johnson1987contact}, we adopt a frame of reference with the axes centered at the midpoint of the width of the contact area and orientated as shown in Figure~\ref{fig:fw}. Forces are taken as positive when pointed towards positive values of the axes direction, the consequent convention for positive values of the stresses being shown in Figure~\ref{fig:fw}.

We then introduce:
\begin{itemize}
\item the effective elastic modulus $E^* =\left( \frac{1-\nu_{c}^2}{E_\mathrm{c}} + \frac{1-\nu_\mathrm{s}^2}{E_\mathrm{s}} \right)^{-1}$, where $E_\mathrm{c}$, $\nu_{c}$ and $E_\mathrm{s}$, $\nu_\mathrm{s}$ are the elastic modulus and Poisson's ratio of the cylindrical particle and the opposing surface respectively;
\item the equivalent curvature $\frac{1}{R} = \frac{1}{R_\mathrm{c}} + \frac{1}{R_\mathrm{s}}$, where $R_\mathrm{c}$ and $R_\mathrm{s}$ are the radius of curvature of the cylindrical particle and the opposing surface respectively; note that $\frac{1}{R} \to \frac{1}{R_\mathrm{c}}$ in our case ($R_\mathrm{s} \to \infty$).
\end{itemize}

Throughout the manuscript, the following non-dimensionalization is then adopted\footnote{ Note that the normalization adopted here differs from that of~\cite{maugis1992adhesion} and~\cite{baney1997cohesive}. There, some lengths are non-dimensionalized by $l$ and some others by $a$, which can be confusing -- hence our choice. Here the non-dimensional parameters are those used by~\cite{johnson2008maugis}. The equations of this section and of Section~\ref{sec:calculation} are presented in the Maugis formalism in \ref{asec:maugs} for the reader who is more familiar with that approach.}:

\begin{subequations} \label{eq:nondim}
  \begin{align}
    \alpha &\vcentcolon= \frac{a}{l} \label{eq:nondim_a} \\
    \beta &\vcentcolon= \frac{b}{l}  \label{eq:nondim_b} \\
    \chi &\vcentcolon= \frac{c}{l}  \label{eq:nondim_c} \\
    \xi &\vcentcolon= \frac{x}{l}  \label{eq:nondim_x} \\
    l &\vcentcolon= \left( \frac{R^2 w}{E^*} \right)^{1/3} \label{eq:nondim_l} \\
    \lambda &\vcentcolon= \frac{\sigma_0}{s}  \label{eq:nondim_lambda} \\
    s &\vcentcolon= \left( \frac{w E^{*2}}{R} \right)^{1/3} \label{eq:nondim_s} \\
    \tilde{P} &\vcentcolon= \frac{P}{P_0} \label{eq:nondim_P} \\
    \tilde{T} &\vcentcolon= \frac{T}{P_0} \label{eq:nondim_T} \\
    P_0 &\vcentcolon= \left( R w^2 E^* \right)^{1/3}  \label{eq:nondim_P0}
  \end{align}
\end{subequations}
with $a$ and $c$ the half-width of contact and interaction respectively, $\sigma_0$ the limiting adhesive stress that the material can sustain, $w$ the surface energy, $P$ and $T$ the normal and tangential force (per unit length) transmitted at the contact interface (the main symbols with their meaning and dimensions are summarized in Table~\ref{tab:nomenclature}).

The particle is assumed to roll towards positive values of $\xi$, and interacts with the opposite surface over the width $2\chi = 2(\alpha+\beta)$, where $\alpha$ is the non-dimensionalized half-width of contact and $\beta$ is the non-dimensionalized half-width of adhesion, i.e. the region where the bodies interact through the adhesive forces even if not directly in contact. When no adhesion is considered, $\beta \to 0$ and $\chi \to \alpha$.

\vspace{\baselineskip}

In the derivation of the stress fields, it is assumed that the contact width is small compared to the dimensions of the bodies and their radius of curvature. These hypotheses imply~\citep{johnson1987contact} that strains are small, both bodies can be regarded as half-spaces, both surfaces are continuous, and contact is non-conforming (i.e.\ the geometries of the surfaces are dissimilar and contact takes place only at a point or line prior any deformation, like for a sphere on a plane).

\paragraph{Transmitted forces}

The normal and tangential components of the non-dimensionalized transmitted force (per unit length) $\tilde{P}$ and $\tilde{T}$ are related to the non-dimensionalized normal and tangential tractions $\tilde{Z}(\xi)$ and $\tilde{X}(\xi)$ by satisfying equilibrium at the interface:

\begin{subequations}\label{eq:forcesnd}
  \begin{align}
    & \tilde{P} = \int_{-\chi}^{\chi} \tilde{Z}(\xi) \textrm{d}\xi  \label{eq:forcesnd_P} \\
    & \tilde{T} = \int_{-\alpha}^{\alpha} \tilde{X}(\xi) \textrm{d}\xi  \label{eq:forcesnd_T}
  \end{align}
\end{subequations}
where $\tilde{Z}(\xi)$ and $\tilde{X}(\xi)$ are the transmitted normal and tangential stress distributions respectively. $\tilde{Z}(\xi)$ and $\tilde{X}(\xi)$ have same magnitude and opposite sign on the cylinder and on the half-space surfaces. Note that as we consider two-dimensional contact, $\tilde{T}$ and $\tilde{X}(\xi)$ are aligned with the $\xi$-axis. Furthermore, it is assumed that the tangential forces can be transmitted only where the two bodies are in direct contact (i.e.\ $\lvert \xi \rvert \le \alpha$), and that the normal forces can be transmitted across the whole interaction width (i.e.\ $\lvert \xi \rvert \le \chi$).
\renewcommand{\thefootnote}{\alph{footnote}}
Under the assumption of Coulomb friction, the following conditions need be satisfied:

\begin{subequations} \label{eq:muforcesnd}
  \begin{align}
    & \lvert \tilde{T} \rvert \le \mu \tilde{P}  \label{eq:muforces_g_nd}\\
    & \lvert \tilde{X}(\xi) \rvert \le \mu \tilde{Z}(\xi) \textrm{ , } \forall \xi \in \left[-\alpha,\alpha\right]  \label{eq:muforces_l_nd}
  \end{align}
\end{subequations}
where $\mu$ is the coefficient of friction. Equations~\ref{eq:muforces_g_nd} and~\ref{eq:muforces_l_nd} require the Coulomb law to be satisfied globally and locally, respectively. When the equal sign holds in Equation~\ref{eq:muforces_g_nd}, slip takes place, while when the equal sign holds in Equation~\ref{eq:muforces_l_nd}, sliding takes place\footnote{ We define slip the relative motion at a point on the contact interface where the two surfaces have different velocities. If slip takes place at all the points on the contact interface, then we refer to it as sliding or complete slip~\citep{johnson1958effect}.}.

\subsection{Adhesive forces: Maugis model} \label{ssec:maugis}

When adhesion is present, tensile stresses normal to the contact interface can be sustained. The two extreme cases are those of the JKR~\citep{johnson1971surface} and the DMT~\citep{derjaguin1975effect} theories, better suitable respectively for the cases of soft materials with high adhesion, and hard materials with low adhesion. Later, \cite{maugis1992adhesion} developed a comprehensive theory of the adhesion of spheres, which is able to capture the continuous transition from the JKR to the DMT limit. In his 3D contact model, Maugis assumes that the part of adhesive forces acting outside the contact area between the sphere and the half-plane are constant and analogue to the cohesive forces between the lips of a crack, as in~\cite{dugdale1960yielding}. These adhesive forces act within an adhesive ring of length $\beta=\chi-\alpha$ (cf. Figure~\ref{fig:fw}). To study the conditions under which the crack propagates (or closes), i.e. the contact radius decreases (or increases), the Griffith criterion for equilibrium $G=w$ is investigated, where $G$ is the energy release rate and $w=2\gamma$ is the fracture energy (i.e.\ the work necessary to separate two surfaces characterized by the same surface energy $\gamma$). By imposing the equilibrium condition, the Maugis parameter\footnote{ The work of \cite{maugis1992adhesion} being in 3D, $\lambda$ has actually a different expression then the one derived from Eqs.~\ref{eq:nondim_l} and~\ref{eq:nondim_lambda}: $\lambda_{3\mathrm{D}} = 2\sigma_0 \cdot \left( \uppi w K^2 / R \right)^{-1/3}$, where $K=4/3 \cdot E/\left(1-\nu^2\right)$. One can see that $\lambda$ and $\lambda_{3\mathrm{D}}$ behave the same (cf. Eqs.~\ref{eq:nondim_l} and~\ref{eq:nondim_lambda}), thus for readability the general notation $\lambda$ is kept throughout the text also when referring to the 3D case.} $\lambda$ allows to investigate the problem as a function of the material parameters. When $\lambda \to \infty$ (soft materials, high adhesion), the adhesive ring vanishes ($\beta\to0$), the inner pressure distribution coincides with the well known JKR limit~\citep{johnson1971surface}, the adhesive forces act only within the contact area and go to infinity at the edges of the contact. If $\lambda \to 0$ (hard materials, low adhesion), the adhesive ring is infinitely large ($\beta\to\infty$), and the DMT limit~\citep{derjaguin1975effect} is recovered instead.


Maugis approach was later adopted in two dimensions, i.e. for cylinders~\citep{chaudhury1996adhesive,baney1997cohesive}, and the distribution of non-dimensionalized normal tractions in such case is

\begin{align} \label{eq:maugis}
  & \tilde{Z}(\xi) =
    \begin{cases}
      \begin{aligned}
        & -\frac{1}{2} \sqrt{ \alpha^2-\xi^2} + \frac{2 \lambda}{\uppi} \arctan{ \sqrt{ \frac{\chi^2-\alpha^2}{\alpha^2-\xi^2} }  } &&\lvert \xi \rvert \le \alpha \\
        & -\lambda && \alpha < \lvert \xi \rvert \le \chi \\
        & 0 &&\lvert \xi \rvert > \chi
      \end{aligned}
    \end{cases}
\end{align}

The non-dimensionalized contact half-width $\alpha$ is determined by solving the load equation (\cite{johnson2008maugis})

\begin{align} \label{eq:load}
    \tilde{P} = \frac{\uppi}{4} \alpha^2 - 2 \lambda \alpha \sqrt{ m^2 -  1 }
\end{align}
where

\begin{subequations}
  \begin{align}
    m &\vcentcolon= \frac{\chi}{\alpha} = \frac{\chi}{\alpha} \cdot \frac{l}{l} = \frac{c}{a}  \label{eq:nondim_m}
  \end{align}
\end{subequations}
Once $\alpha$ is known as a function of $m$, $m$ is determined by solving the equation for the cut-off distance at which the adhesive stresses fall to zero:

\begin{align}
    \lambda \alpha^2 g_1(m) + \lambda^2 \alpha g_2(m) = 1
\end{align}
where

\begin{align}
  g_1(m) &\vcentcolon= \frac{1}{2} \left[ m \sqrt{ m^2 - 1 } - \cosh^{-1}\left(m\right) \right] \\
  g_2(m) &\vcentcolon= \frac{4}{\uppi} \left[ \cosh^{-1}\left(m\right) \sqrt{ m^2 - 1 } - m \ln\left(m\right) \right]
\end{align}
and the values of $\alpha$ and $\chi$ are eventually found. Following the aforementioned steps, it is possible to fully determine the non-dimensionalized tractions $\tilde{Z}(\xi)$ at the interface. For the details of the derivation we refer the reader to~\cite{maugis1992adhesion,baney1997cohesive,johnson2008maugis,wu2009adhesive}.


\subsection{Tractive rolling forces - Carter model} \label{ssec:carter}

To the best of our knowledge, no analytical solution for tractive rolling contact in the presence of adhesive forces have been found yet. Here, we assume that it can be obtained by superposition of the analytical solutions for the normal adhesive contact (described above) and the classical solution for the case of a cylinder rolling on an elastic half-space (described in the following). Contact then takes place along a frictional interface (characterized by the friction coefficient $\mu$) and the tangential forces due to the rolling motion are transmitted between the two bodies along the contact area ($\lvert \xi \rvert \le \alpha$). The problem was first solved by~\cite{carter1926action}, with an approach that is conceptually similar to that of~\cite{cattaneo1938sul} for the contact of cylinders with partial slip -- two tangential distributions of the Hertzian type are superimposed, so that Coulomb law (Equation ~\ref{eq:muforces_l_nd}) is satisfied everywhere along the contact area.


The two distributions, non-dimensionalized, are

\begin{subequations} \label{eq:trac_carter}
\begin{align}
 & \tilde{X}'(\xi) =   \begin{cases}
    \begin{aligned}
    & \mu \frac{2\tilde{P}}{\uppi \alpha^2} \sqrt{ \alpha^2 - \xi^2 } &&\lvert \xi \rvert \le \alpha \\
    & 0 &&\lvert \xi \rvert > \alpha
    \end{aligned}
  \end{cases} \label{eq:trna_Xp}\\
 & \tilde{X}''(\xi) = \begin{cases}
    \begin{aligned}
    & - \mu \frac{2\tilde{P}}{\uppi \alpha^2} \sqrt{ \rho^2 - \left(\xi-\psi\right)^2 } &&\lvert \xi-\psi \rvert \le \rho \\
    & 0 &&\lvert \xi - \psi \rvert > \rho
    \end{aligned}
  \end{cases} \label{eq:X}
\end{align}
\end{subequations}
where $\rho \vcentcolon = \alpha - \psi = (1-\frac{T}{\mu P})^{1/2} \le 1$ is the non-dimensionalized half-width over which $\tilde{X}''(\xi)$ is applied. Note that $\tilde{X}'(\xi)=\mu \tilde{Z}(\xi)$ corresponds to the tangential distribution in the case of sliding, and the superposition of $\tilde{X}''(\xi)$ allows to take into account the stick region~\citep{johnson1958effect}. The total tangential tractions at the interface are then (\cite{johnson1958effect}, see Figure~\ref{fig:tangential_load}a)

\begin{align} \label{eq:trna_X}
  \tilde{X}(\xi) = \tilde{X}'(\xi) + \tilde{X}''(\xi)
\end{align}
and are positive when transmitted from the particle to the surface, to oppose the direction of motion. The Coulomb law (Equations~\ref{eq:muforcesnd}) is satisfied both globally and locally.


\subsection{Stress field}\label{ssec:stresses}

We now assume that the total Cauchy stress $\sigma_{ij}(x,y,z)$, with $i,j=x,y,z$, is made up of the linear superposition of two different contributions: $\sigma_{ij}^\mathrm{n}$ due to the normal component of the transmitted force, and $\sigma_{ij}^\mathrm{t}$ due to the tangential component of the transmitted force. We define then the total non-dimensionalized stress $\bar{ \tilde{\sigma}}_{ij}(\xi)$ anywhere along the surface as

\begin{equation} \label{eq:sigma_red}
  \bar{ \tilde{\sigma}}_{ij}(\xi) = \bar{ \tilde{\sigma} }_{ij}^\mathrm{n}(\xi) + \bar{ \tilde{\sigma} }_{ij}^\mathrm{t}(\xi)
\end{equation}
where the overbar indicates that the stresses are considered at $\zeta=0$. The assumption of plain strains reduces the number of independent components of stress, and in Eq.~\ref{eq:sigma_red} is $\bar{\tilde{\sigma}}_{yy}(\xi) = \nu \left( \bar{\tilde{\sigma}}_{xx}(\xi) + \bar{\tilde{\sigma}}_{zz}(\xi) \right)$. We will then not consider explicitly $\bar{\tilde{\sigma}}_{yy}(\xi)$ in the rest of the manuscript.

Furthermore, to satisfy equilibrium and assuming that the tractions $\tilde{Z}(\xi)$ and $\tilde{X}(\xi)$ are specified independently, it is

\begin{equation} \label{eq:sigma_red_eq}
\begin{aligned}
  & \bar{ \tilde{\sigma}}_{zz}(\xi) = \bar{ \tilde{\sigma}}_{zz}^\mathrm{n}(\xi) = - \tilde{Z}(\xi) \\
  & \bar{ \tilde{\sigma}}_{xz}(\xi) = \bar{ \tilde{\sigma}}_{xz}^\mathrm{t}(\xi) = - \tilde{X}(\xi) \quad \textrm{.}
\end{aligned}
\end{equation}
and thus the stress state of Equation~\ref{eq:sigma_red} reduces to

\begin{equation} \label{eq:sigma_red_cmp}
\begin{aligned}
  & \bar{ \tilde{\sigma}}_{xx}(\xi) = \bar{ \tilde{\sigma}}_{xx}^\mathrm{n}(\xi) + \bar{ \tilde{\sigma}}_{xx}^\mathrm{t}(\xi) \\
  & \bar{ \tilde{\sigma}}_{zz}(\xi) = \bar{ \tilde{\sigma}}_{zz}^\mathrm{n}(\xi) \\
  & \bar{ \tilde{\sigma}}_{xz}(\xi) = \bar{ \tilde{\sigma}}_{xz}^\mathrm{t}(\xi) \quad \textrm{.}
\end{aligned}
\end{equation}

In the absence of tangential tractions at the interface, the plain strain assumption in elasticity also implies that~\citep{johnson1987contact}

\begin{equation}\label{eq:sxsz}
  \bar{ \tilde{\sigma}}_{xx}(\xi) =  \bar{ \tilde{\sigma}}_{zz}(\xi) = -\tilde{Z}(\xi) \quad \textrm{.}
\end{equation}

\paragraph{Stresses due to normal load}

\begin{figure*}
  \centering
  \begin{subfigure}{.32\textwidth}
    \centering
    \includegraphics[width=\textwidth]{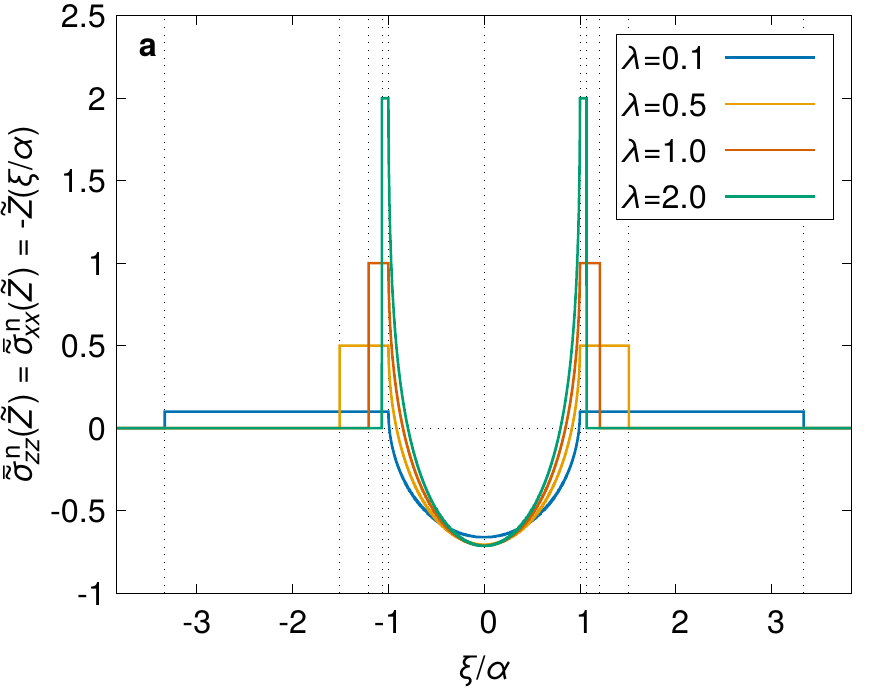}
  \end{subfigure}
  \begin{subfigure}{.32\textwidth}
    \centering
    \includegraphics[width=\textwidth]{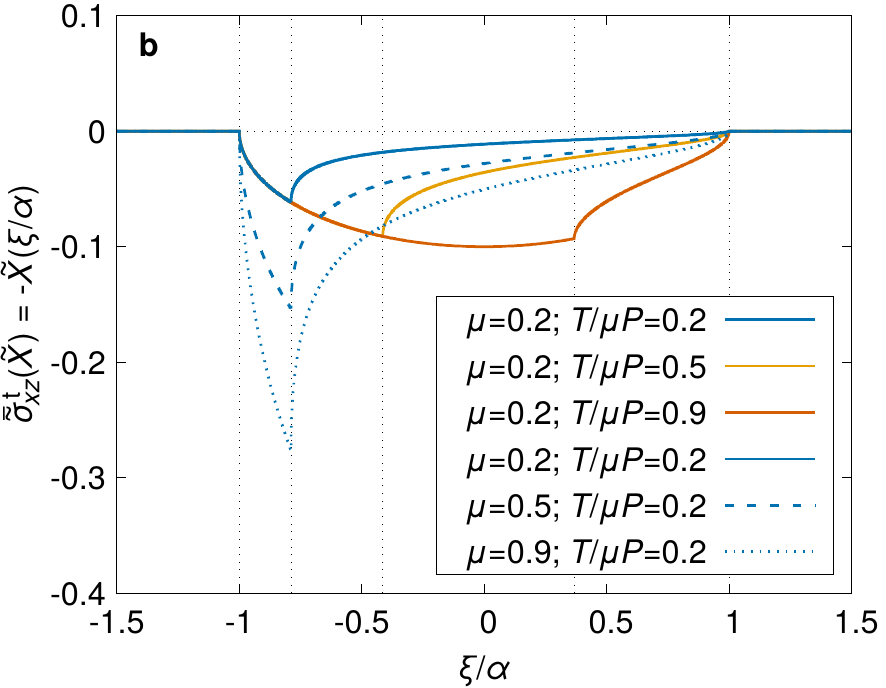}
  \end{subfigure}
  \begin{subfigure}{.32\textwidth}
    \centering
    \includegraphics[width=\textwidth]{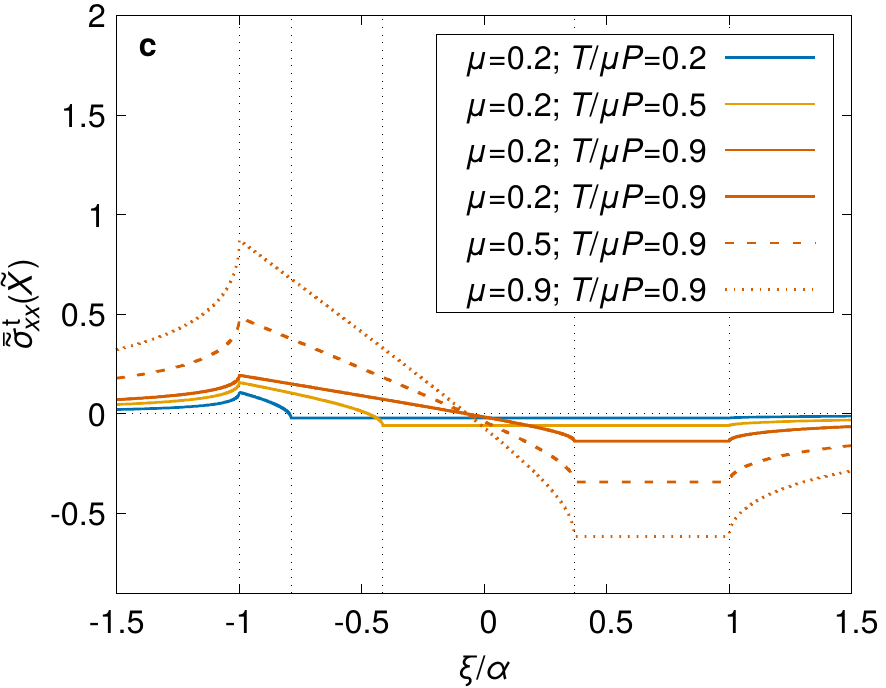}
  \end{subfigure}
  \caption{Tractions and stress distributions. The horizontal axis is scaled by $\alpha$. a) Stresses $\bar{ \tilde{\sigma}}_{ij}^\mathrm{n}(\xi)$ due to normal load of the Maugis type (Equations~\ref{eq:trna_eq}); larger $\lambda$ values give larger tensile stresses at the edge of contact and smaller adhesive width. b) Tangential stress $\bar{ \tilde{\sigma}}_{xz}^\mathrm{t}(\xi)$ due to tangential load (Equation~\ref{eq:carter_xz}); $T / \mu P$ affects the shape of the distribution (the larger the ratio, the larger the slip zone), $\mu$ scales it. c) Normal stress $\bar{ \tilde{\sigma}}_{xx}^\mathrm{t}(\xi)$ due to tangential load (Equation~\ref{eq:carter_xx}); $T / \mu P$ affects the shape of the distribution, $\mu$ scales it. In all the figures $\tilde{P}=\uppi / 2$.}
  \label{fig:tractions}
\end{figure*}

From Equations~\ref{eq:maugis},~\ref{eq:sigma_red_cmp} and~\ref{eq:sxsz}, it follows that the surface stresses due to the normal component of the force transmitted at the interface are (see Figure~\ref{fig:tractions}a)

\begin{equation} \label{eq:trna_eq}
\begin{aligned}
  & \bar{ \tilde{\sigma}}_{xx}^\mathrm{n}(\xi) = \bar{ \tilde{\sigma}}_{zz}^\mathrm{n}(\xi) =
    \begin{cases}
    \begin{aligned}
    & \frac{1}{2} \sqrt{ \alpha^2-\xi^2} - \frac{2 \lambda}{\uppi} \arctan{ \sqrt{ \frac{\chi^2-\alpha^2}{\alpha^2-\xi^2} }  } &&\lvert \xi \rvert \le \alpha \\
    & \lambda && \alpha < \lvert \xi \rvert \le \chi \\
    & 0 &&\lvert \xi \rvert > \chi
    \end{aligned}
  \end{cases} \\
  & \bar{ \tilde{\sigma}}_{xz}^\mathrm{n}(\xi) = 0
\end{aligned} \quad \textrm{.}
\end{equation}

To fully determine $\bar{ \tilde{\sigma}}_{ij}^\mathrm{n}(\xi)$ is thus necessary and sufficient to know the non-dimensionalized normal component $\tilde{P}$ of the transmitted force and the Maugis parameter $\lambda$.

In Equations~\ref{eq:maugis} and ~\ref{eq:trna_eq}, no discontinuity is present between the constant value $\lambda$ of the non-dimensionalized adhesive forces and the pressure distribution within the contact width, i.e.\ $\lim_{\lvert \xi \rvert \to \alpha^\pm } \bar{\tilde{\sigma}}_{zz} = \lambda$, and compressive pressures are described at the center and tensile pressures at the edge of the contact. A discontinuity in the normal stresses is present instead at the edges of the interaction area, where $\lim_{\xi \to -\chi^+ } \bar{\tilde{\sigma}}_{zz} = \lambda \neq \lim_{\xi \to -\chi^- } \bar{\tilde{\sigma}}_{zz} = 0$ (and similarly for $\xi \to \chi^-$).

\paragraph{Stresses due to tangential load}

The stresses due to the transmitted tangential tractions $\tilde{X}(\xi)$ are determined from the superposition of the stresses due to $\tilde{X}'(\xi)$ and $\tilde{X}''(\xi)$, which have the same shape but different magnitude and opposite sign.

\begin{figure*}
  \centering
  \begin{subfigure}{.49\textwidth}
    \centering
    \includegraphics[width=\textwidth]{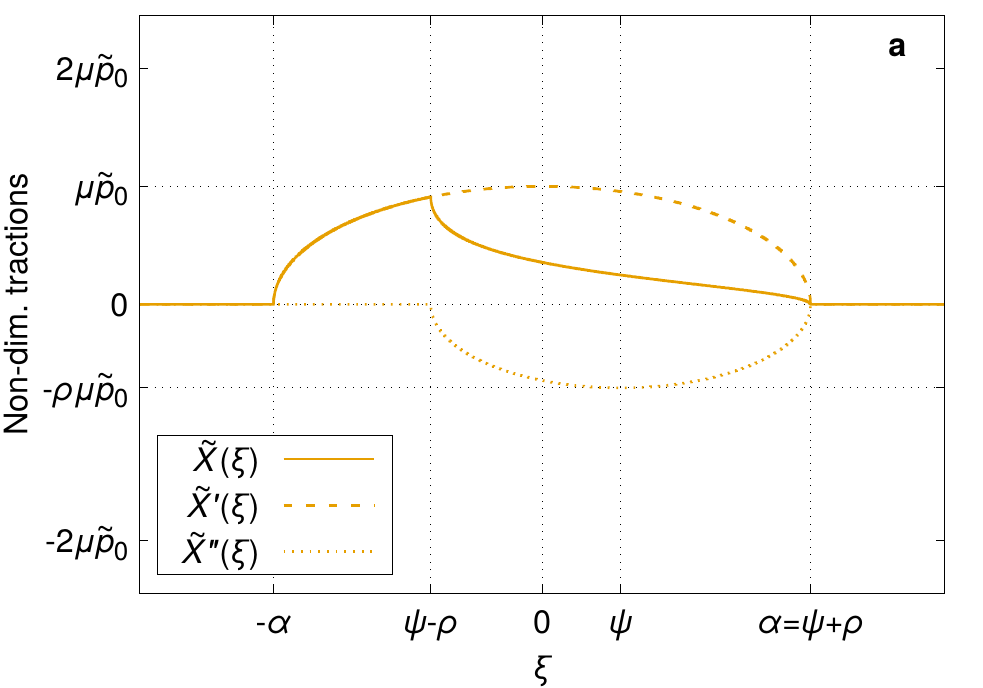}
  \end{subfigure}
  \begin{subfigure}{.49\textwidth}
    \centering
    \includegraphics[width=\textwidth]{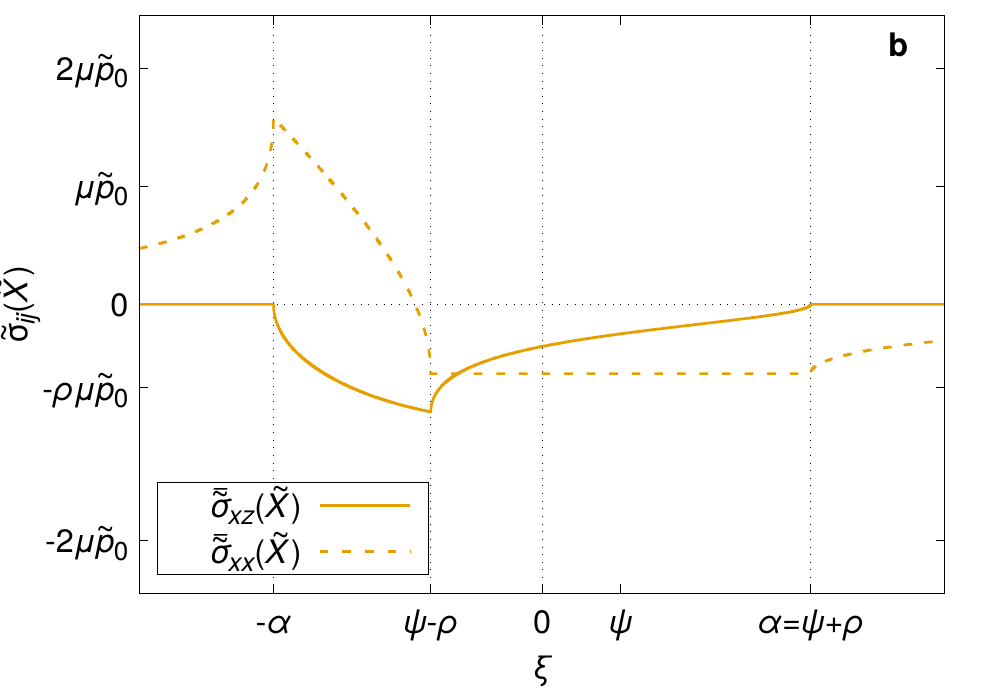}
  \end{subfigure}
  \caption{Construction of the tangential tractions and relative stresses. a) Transmitted tangential tractions $\tilde{X}(\xi)=\tilde{X}'(\xi)+\tilde{X}''(\xi)$ (Equations~\ref{eq:carter}); slip takes place over the strip $-\alpha \le \xi \le \psi-\rho$, whose size increases with $T / \mu P$ (cf. Figure~\ref{fig:tractions}) and where is $\lvert \tilde{X}(\xi) \rvert=\mu \lvert \tilde{Z}(\xi) \rvert$. b) Stresses $\bar{ \tilde{\sigma}}_{ij}^\mathrm{t}(\xi)$ due to the tractions $\tilde{X}(\xi)$. $\pm 2 \mu \tilde{p}_0$ are the limiting values for $\bar{ \tilde{\sigma}}_{xx}^\mathrm{t}(\xi \to \mp \alpha)$ and they are reached when $T / \mu P \to 1$. In both figures $\tilde{P}=\uppi / 4$ and thus $\alpha=1$ (cf. Equation~\ref{eq:load}).}
  \label{fig:tangential_load}
\end{figure*}

The surface stresses due to the tangential component of force transmitted at the interface are then (see Figure~\ref{fig:tangential_load}b)

\begin{subequations} \label{eq:carter}
\begin{align}
  & \bar{ \tilde{\sigma}}_{zz}^\mathrm{t}(\xi) = 0  \label{eq:carter_zz}\\
  \begin{split} \label{eq:carter_xz}
    & \bar{ \tilde{\sigma}}_{xz}^\mathrm{t}(\xi) = - \tilde{X}(\xi) = - \tilde{X}'(\xi) - \tilde{X}''(\xi) = \\
    &=\begin{cases}
      \begin{aligned}
        & -\mu \frac{2\tilde{P}}{\uppi \alpha^2} \sqrt{ \alpha^2 - \xi^2 } &&\lvert \xi \rvert \le \alpha  \\
        & 0 &&\lvert \xi \rvert > \alpha
      \end{aligned}
    \end{cases} +
    \begin{cases}
      \begin{aligned}
        & \mu \frac{2\tilde{P}}{\uppi \alpha^2} \sqrt{ \rho^2 - \left(\xi-\psi\right)^2 } &&\lvert \xi-\psi \rvert \le \rho \\
        & 0 &&\lvert \xi - \psi \rvert > \rho
      \end{aligned}
    \end{cases}
  \end{split}\\
  \begin{split}  \label{eq:carter_xx}
    &\bar{ \tilde{\sigma}}_{xx}^\mathrm{t}(\xi) =  \bar{ \tilde{\sigma}}_{xx}^{\tilde{X}'}(\xi) + \bar{ \tilde{\sigma}}_{xx}^{\tilde{X}''}(\xi) = \\
    &=\begin{cases}
      \begin{aligned}
        & -2 \mu \frac{2\tilde{P}}{\uppi \alpha^2} \xi &&\lvert \xi \rvert \le \alpha \\
        & -2 \mu \frac{2\tilde{P}}{\uppi \alpha^2} \left( \xi - \sgn\left( \xi \right) \sqrt{ \xi^2-\alpha^2 } \right) &&\lvert \xi \rvert > \alpha
      \end{aligned}
    \end{cases} +\\
    &+\begin{cases}
      \begin{aligned}
        & 2 \mu \frac{2\tilde{P}}{\uppi \alpha^2} (\xi - \psi) &&\lvert \xi-\psi \rvert \le \rho \\
        & 2 \mu \frac{2\tilde{P}}{\uppi \alpha^2} \left[ \xi - \psi - \sgn\left( \xi - \psi \right) \sqrt{ \left(\xi -\psi\right)^2 - \rho^2 }\right] &&\lvert \xi-\psi \rvert > \rho
      \end{aligned}
    \end{cases}
    \end{split}
\end{align}
\end{subequations}
and the effects of $\mu$ and $T/\mu P$ are depicted in Figures~\ref{fig:tractions}a and~\ref{fig:tractions}b. For the derivation of Equation~\ref{eq:carter_xx} we refer the reader to~\cite{johnson1987contact}.

\subsection{Crack propagation angle}\label{ssec:cangle}

Once the stress field is determined, the principal stresses are then computed as:

\begin{align} \label{eq:sigma_pr}
  & \bar{ \tilde{\sigma}}_1(\xi) = \frac{\bar{ \tilde{\sigma} }_{xx}(\xi) + \bar{ \tilde{\sigma} }_{zz}(\xi)}{2} + \sqrt{ \left( \frac{\bar{ \tilde{\sigma} }_{xx}(\xi) - \bar{ \tilde{\sigma} }_{zz}(\xi)}{2} \right)^2 + \bar{ \tilde{\sigma} }_{xz}^2(\xi) } \\
  & \bar{ \tilde{\sigma}}_2(\xi) = \frac{\bar{ \tilde{\sigma} }_{xx}(\xi) + \bar{ \tilde{\sigma} }_{zz}(\xi)}{2} - \sqrt{ \left( \frac{\bar{ \tilde{\sigma} }_{xx}(\xi) - \bar{ \tilde{\sigma} }_{zz}(\xi)}{2} \right)^2 + \bar{ \tilde{\sigma} }_{xz}^2(\xi) }
\end{align}
such that $\bar{ \tilde{\sigma}}_1(\xi) \ge \bar{ \tilde{\sigma}}_2(\xi)$ for all $\xi$.

\begin{figure*}
  \centering
  \includegraphics[width=\textwidth]{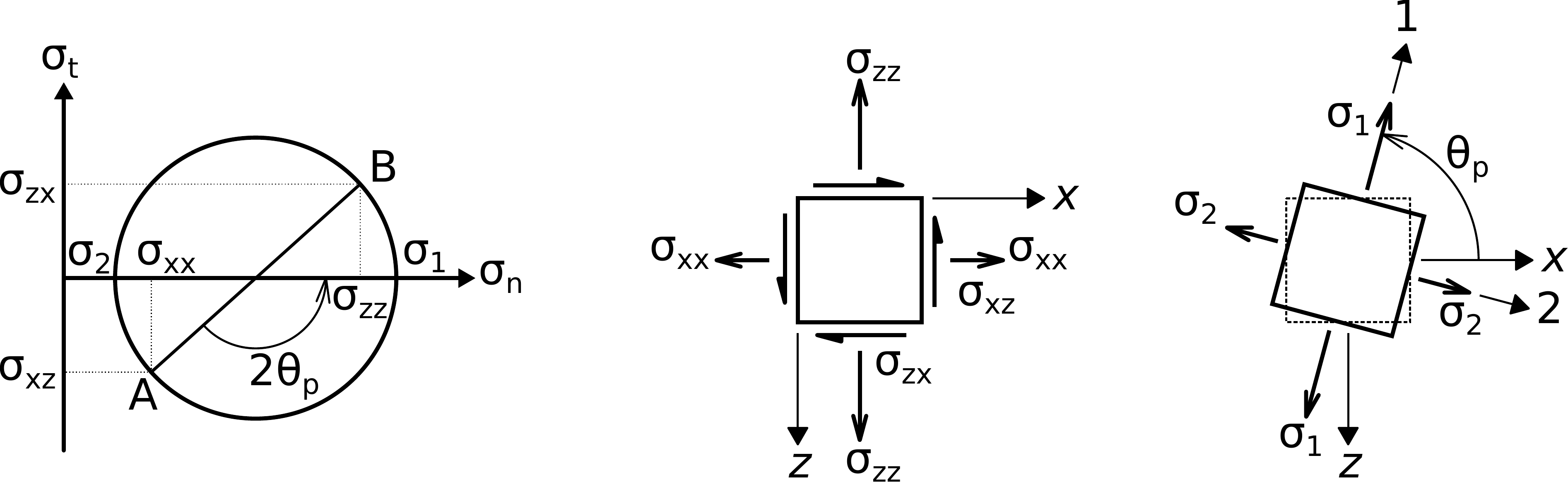}
  \caption{Mohr diagram (left) for a generic stress state (centre) and principal stresses directions (right). According to Mohr's convention, shear stresses are positive when rotating the representative square clockwise.}
  \label{fig:mohr}
\end{figure*}

The stress field $\bar {\tilde{\sigma}}_{ij}(\xi,\eta,\zeta)$ at a given point can be visualized in the Mohr plane (Figure~\ref{fig:mohr}), following the convention that shear stresses are positive when rotating the representative square clockwise. This implies that $\bar{\tilde{\sigma}}_{xz}$ has opposite sign in the Mohr plane with respect to the contact mechanics convention (cf. Section~\ref{ssec:fw} and Figure~\ref{fig:fw}). Note that for the problem at hand $\bar{\tilde{\sigma}}_{xy}=\bar{\tilde{\sigma}}_{yz}=0$ (because of symmetry) and $\eta$ is the principal direction of the principal stress $\bar{\tilde{\sigma}}_3 \equiv \bar{ \tilde{\sigma}}_{yy} \textrm{ } \forall \xi$. The other two principal directions lie then on the $<\xi,\zeta>$ plane and they can be determined by analysing only the stresses $\bar{\tilde{\sigma}}_{xx}$, $\bar{\tilde{\sigma}}_{zz}$, and $\bar{\tilde{\sigma}}_{xz}$\footnote{ While we focus here only on the surface stress $\bar {\tilde{\sigma}}_{ij}(\xi)$, these arguments hold for the general stress state ${\tilde{\sigma}}_{ij}(\xi,\eta,\zeta)$ in the problem at hand, i.e.\ $\tilde{\sigma}_{xy}=\tilde{\sigma}_{yz}=0$ always and $\tilde{\sigma}_3 \equiv \tilde{\sigma}_{yy} \textrm{ } \forall (\xi,\eta,\zeta)$, $\eta$ is principal direction, and the other two directions always lie on the $<\xi,\zeta>$ plane.}.

The angle $\theta_\mathrm{p}$ between the direction of the largest principal stress $\bar{ \tilde{\sigma}}_1(\xi)$ and the $\xi$-axis (cf. Figure~\ref{fig:fw}) is then

\begin{equation}\label{eq:thetap}
  \begin{aligned}
    \theta_\mathrm{p}(\xi) = \begin{cases}
      -\frac{\sgn(\bar{ \tilde{\sigma} }_{xz}(\xi))}{2} \cdot \arctan \left( \frac{2 \lvert \bar{ \tilde{\sigma} }_{xz}(\xi) \rvert}{\bar{ \tilde{\sigma} }_{xx}(\xi) - \bar{ \tilde{\sigma} }_{zz}(\xi)} \right) & \bar{ \tilde{\sigma} }_{xx}(\xi) \ge \bar{ \tilde{\sigma} }_{zz}(\xi) \\
      -\frac{\sgn(\bar{ \tilde{\sigma} }_{xz}(\xi))}{2} \cdot \left[ \uppi + \arctan \left( \frac{2 \lvert \bar{ \tilde{\sigma} }_{xz}(\xi) \rvert}{\bar{ \tilde{\sigma} }_{xx}(\xi) - \bar{ \tilde{\sigma} }_{zz}(\xi)} \right) \right] & \bar{ \tilde{\sigma} }_{xx}(\xi) < \bar{ \tilde{\sigma} }_{zz}(\xi)
    \end{cases} \quad \textrm{.}
  \end{aligned}
\end{equation}

Hence the angle $\theta_\mathrm{c}$ between the direction of crack propagation and the $\xi$-axis (cf. Figure~\ref{fig:fw}) is

\begin{equation}\label{eq:thetac}
  \theta_\mathrm{c}(\xi) = \theta_\mathrm{p}(\xi) - \frac{\uppi}{2} \quad \textrm{.}
\end{equation}

In the next sections and in \ref{sec:app1} we will see how it is always the case that $\bar {\tilde{\sigma}}_{xy}(\xi \to -\chi^+) \le 0$, and thus $-\frac{\uppi}{2} \le \theta_\mathrm{c} \le 0$: crack propagation within the half-space is favoured over crack propagation inside the particle.

Finally, we assume that the rolling motion is equivalent to applying the following infinitesimal stresses ($\delta \tilde{\sigma}$ being positive):

\begin{equation}\label{eq:sigma_roll}
  \begin{cases}
    \begin{aligned}
      \delta \tilde{\sigma} &> 0 \quad \textrm{if} \quad -\chi \le \xi \le -\alpha \\
      -\delta \tilde{\sigma} &< 0 \quad \textrm{if} \quad \quad \alpha \le \xi \le \chi
    \end{aligned}
  \end{cases}
\end{equation}
According to the fracture mechanics picture of contact introduced by~\cite{maugis1992adhesion}, this leads to closing the crack at the leading edge and opening it at the trailing edge --- the particle now rolls towards positive values of $\xi$.


\section{Calculation}\label{sec:calculation}

In the present section we present the surface stresses from Section~\ref{sec:theory} for the most general case only: tractive rolling with adhesion.  A detailed discussion on the crack propagation angle and on the influence of the various parameters is reported in Section~\ref{sec:results}. Three more cases, that is free rolling with and without adhesion and tractive rolling without adhesion, are reported and discussed in \ref{sec:app1}.

\begin{figure*}
  \centering
  \begin{subfigure}{.32\textwidth}
    \centering
    \includegraphics[width=\textwidth]{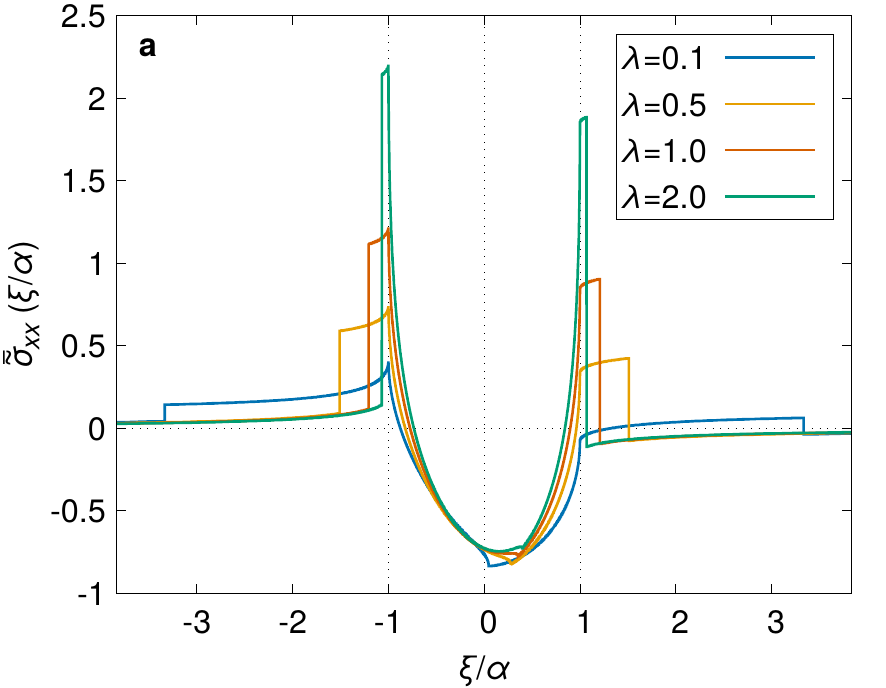}
  \end{subfigure}
  \begin{subfigure}{.32\textwidth}
    \centering
    \includegraphics[width=\textwidth]{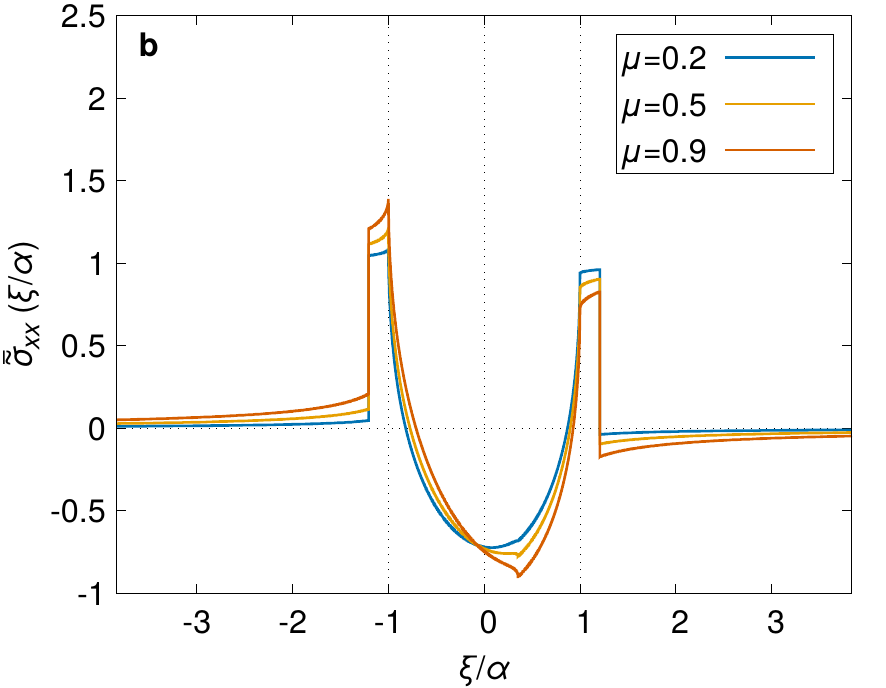}
  \end{subfigure}
  \begin{subfigure}{.32\textwidth}
    \centering
    \includegraphics[width=\textwidth]{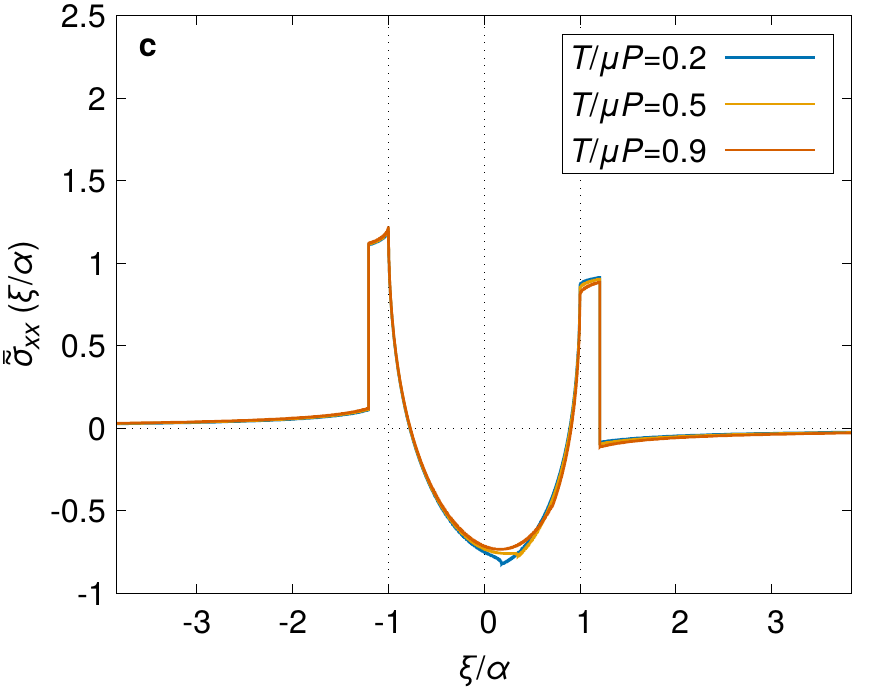}
  \end{subfigure}
  \caption{Stress $\bar{ \tilde{\sigma}}_{xx}(\xi)$ in the case of tractive rolling with adhesion. The horizontal axis is scaled by $\alpha$. a) Effect of the Maugis parameter $\lambda$ for constant values of the friction coefficient and of the ratio between the transmitted tangential force and the limiting frictional force (here: $\mu=0.5$ and $T / \mu P=0.5$). b) Effect of the friction coefficient $\mu$ for constant values of the Maugis parameter and of the ratio between the transmitted tangential force and the limiting frictional force (here: $\lambda=1$ and $T / \mu P=0.5$). c) Effect of the ratio  $T / \mu P$ between the transmitted tangential force and the limiting frictional force for constant values of the Maugis parameter and of the coefficient of friction (here: $\lambda=1$ and $\mu=0.5$). In all the figures $\tilde{P}=\uppi / 4$.}
  \label{fig:trwa_xx}
\end{figure*}

When adhesion is present and the transmitted force has a tangential component, the stress state is given by Equations~\ref{eq:sigma_red_cmp}, with $\bar{\tilde{\sigma}}_{ij}^\mathrm{n}(\xi)$ given by Equations~\ref{eq:trna_eq} and with $\bar{\tilde{\sigma}}_{ij}^\mathrm{t}(\xi)$ given by Equations~\ref{eq:carter}:

\begin{equation}\label{eq:trwa_sigma}
  \begin{aligned}
    &\bar{ \tilde{\sigma}}_{zz}(\xi) =
    \begin{cases}
    \frac{\tilde{p}_0}{\alpha} \sqrt{ \alpha^2-\xi^2} - \frac{2 \lambda}{\uppi} \arctan{ \sqrt{ \frac{\chi^2-\alpha^2}{\alpha^2-\xi^2} } } &\lvert \xi \rvert \le \alpha \\
    \lambda & \alpha < \lvert \xi \rvert \le \chi \\
    0 &\lvert \xi \rvert > \chi
  \end{cases} \\
    &\bar{ \tilde{\sigma}}_{xx}(\xi) =
    \begin{cases}
        -\frac{\tilde{p}_0}{\alpha} \left( \sqrt{ \alpha^2 - \xi^2 } -2 \mu \xi \right)- \frac{2 \lambda}{\uppi} \arctan{ \sqrt{ \frac{\chi^2-\alpha^2}{\alpha^2-\xi^2} } } &\lvert \xi \rvert \le \alpha \\
        -2 \mu \frac{\tilde{p}_0}{\alpha} \left( \xi - \sgn\left( \xi \right) \sqrt{ \xi^2-\alpha^2 } \right) + \lambda &\alpha < \lvert \xi \rvert \le \chi  \\
        -2 \mu \frac{\tilde{p}_0}{\alpha} \left( \xi - \sgn\left( \xi \right) \sqrt{ \xi^2-\alpha^2 } \right) &\lvert \xi \rvert > \chi
    \end{cases} +\\
    &\quad \quad \quad + \begin{cases}
        2 \mu \frac{\tilde{p}_0}{\alpha} (\xi - \psi) &\lvert \xi-\psi \rvert \le \rho \\
        2 \mu \frac{\tilde{p}_0}{\alpha} \left[ \xi - \psi - \sgn\left( \xi - \psi \right) \sqrt{ \left(\xi -\psi\right)^2 - \rho^2 }\right] &\lvert \xi-\psi \rvert > \rho
    \end{cases}\\
  &\bar{ \tilde{\sigma}}_{xz}(\xi) =
    \begin{cases}
        -\mu \frac{\tilde{p}_0}{\alpha} \sqrt{ \alpha^2 - \xi^2 } &\lvert \xi \rvert \le \alpha \\
        0 &\lvert \xi \rvert > \alpha
    \end{cases} +
    \begin{cases}
        \mu \frac{\tilde{p}_0}{\alpha} \sqrt{ \rho^2 - \left(\xi-\psi\right)^2 } &\lvert \xi-\psi \rvert \le \rho \\
        0 &\lvert \xi - \psi \rvert > \rho
    \end{cases}
  \end{aligned}
\end{equation}
where $\tilde{p}_0 \vcentcolon= \frac{2 \tilde{P}}{ \uppi \alpha} = \frac{\alpha}{2}$ is the pre-factor written in the classical Hertz convention (cf. Equation~\ref{eq:load}, with $m \to 0$) and represents the maximum value of the tractions at the center of the contact width. The distributions of $\bar{ \tilde{\sigma}}_{zz}(\xi)$ and $\bar{ \tilde{\sigma}}_{xz}(\xi)$ are depicted in Figures~\ref{fig:tractions}a and~\ref{fig:tractions}b, $\bar{ \tilde{\sigma}}_{xx}(\xi)$ in Figure~\ref{fig:trwa_xx}. Note that the non-dimensionalized load $\tilde{P}$, the friction coefficient $\mu$ and the ratio $T/\mu P$ need to be known to fully determine the stress field $\bar{ \tilde{\sigma}}_{ij}(\xi)$.

\section{Results and discussion}\label{sec:results}
In this section we examine the principal stresses that arise from the stress field derived in Section~\ref{sec:calculation} and their principal direction. From that, we can determine the direction of crack propagation (cf. Equations~\ref{eq:thetap} and~\ref{eq:thetac}). We will finally discuss the effects of different geometries and material parameters on the crack propagation angle.

\begin{figure*}
  \centering
  \begin{subfigure}{.49\textwidth}
    \centering
    \includegraphics[width=\textwidth]{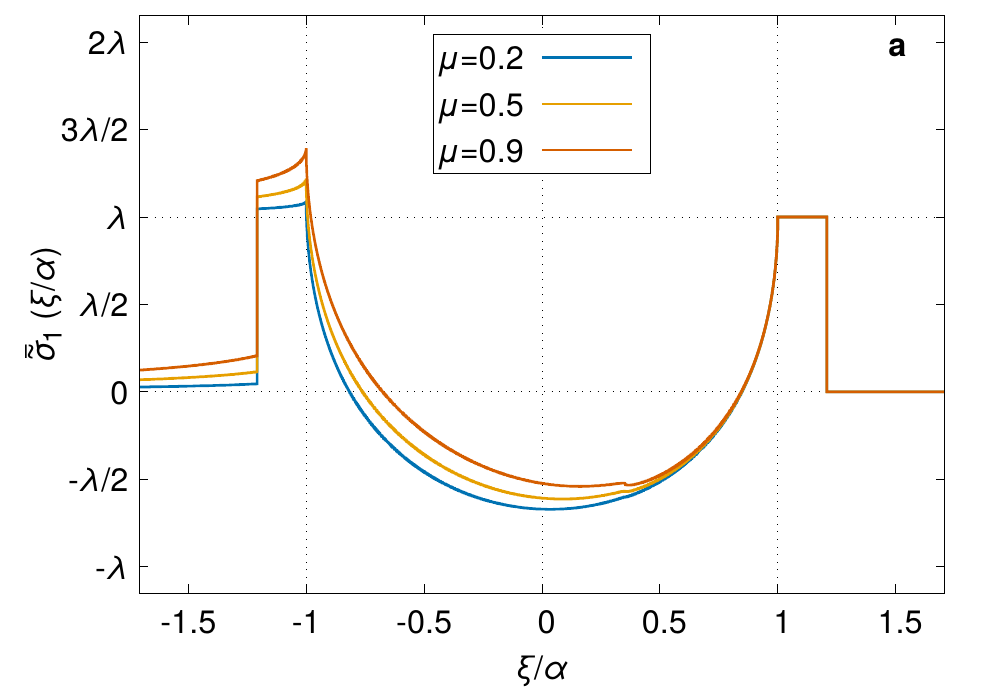}
  \end{subfigure}
  \begin{subfigure}{.49\textwidth}
    \centering
    \includegraphics[width=\textwidth]{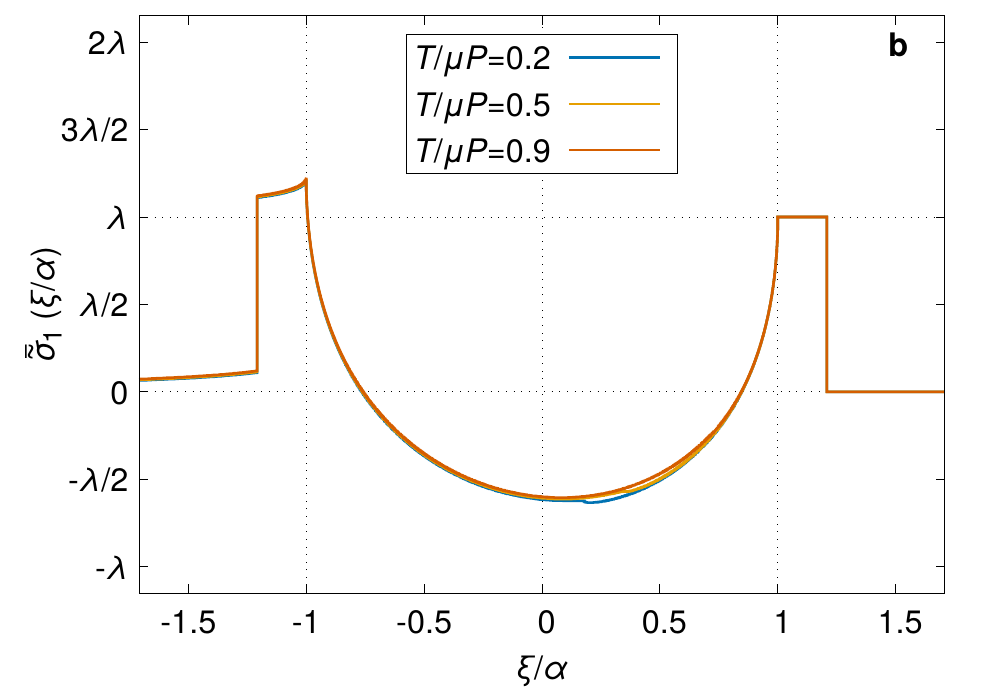}
  \end{subfigure}
  \begin{subfigure}{.49\textwidth}
    \centering
    \includegraphics[width=\textwidth]{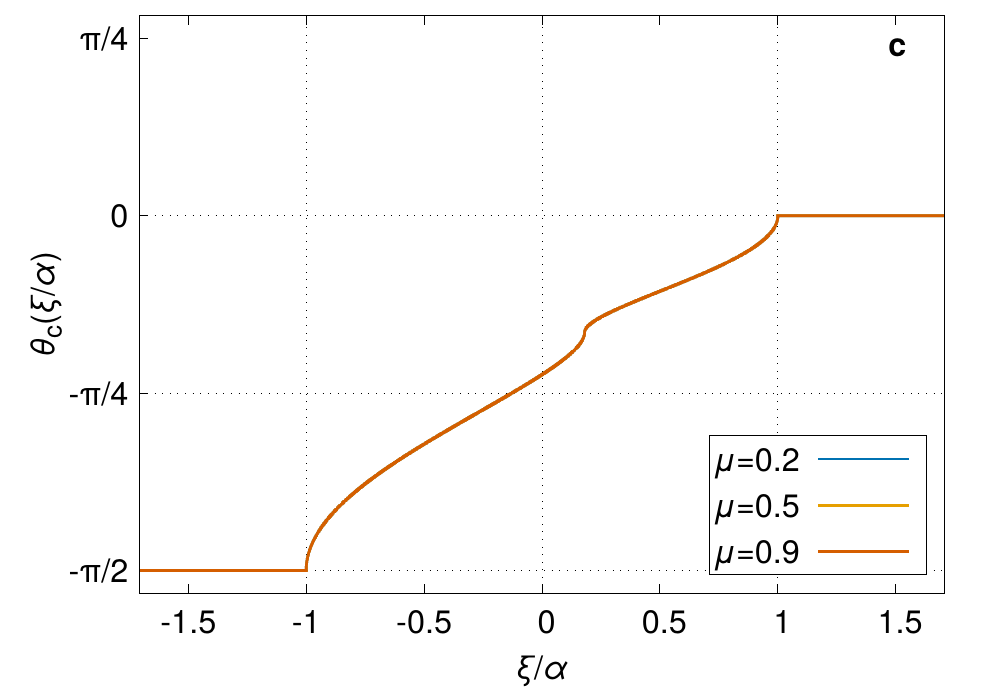}
  \end{subfigure}
  \begin{subfigure}{.49\textwidth}
    \centering
    \includegraphics[width=\textwidth]{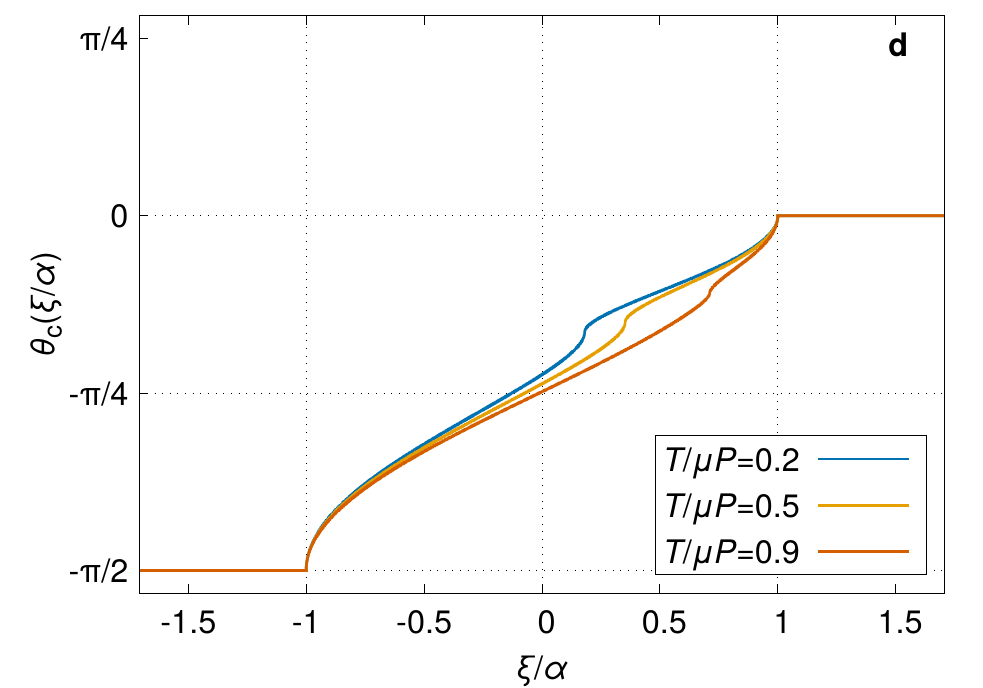}
  \end{subfigure}
  \caption{Maximum principal stress $\bar{ \tilde{\sigma}}_1(\xi)$ and crack propagation angle $\theta_\mathrm{c}$ along the interface in the case of tractive rolling with adhesion. The horizontal axis is scaled by $\alpha$. a) Effect of the friction coefficient $\mu$ on $\bar{ \tilde{\sigma}}_1(\xi)$ for constant values of the Maugis parameter and of the ratio between the transmitted tangential force and the limiting frictional force (here: $\lambda=1$ and $T / \mu P=0.5$). b) Effect of the ratio  $T / \mu P$ on $\bar{ \tilde{\sigma}}_1(\xi)$ for constant values of the Maugis parameter and of the coefficient of friction (here: $\lambda=1$ and $\mu=0.5$). c) The friction coefficient $\mu$ has no effect on $\theta_\mathrm{c}$ (all the curves superpose), for constant values of the Maugis parameter and of the ratio between the transmitted tangential force and the limiting frictional force (here: $\lambda=1$ and $T / \mu P=0.5$). d) Effect of the ratio  $T / \mu P$ on $\theta_\mathrm{c}$ for constant values of the Maugis parameter and of the coefficient of friction (here: $\lambda=1$ and $\mu=0.5$).}
  \label{fig:rtrwa}
\end{figure*}

Figures~\ref{fig:rtrwa}a and~\ref{fig:rtrwa}b show the maximum principal stress $\bar{ \tilde{\sigma}}_1(\xi)$ due to the stress field derived in section~\ref{sec:calculation} (Eqs.~\ref{eq:trwa_sigma}). The stress is tensile in a region around both the leading and trailing edge, its maximum being at the trailing edge ($\xi=-\alpha$). If $\bar{ \tilde{\sigma}}_1(-\alpha) \ge \tilde{\sigma}_\mathrm{c}$, where $\tilde{\sigma}_\mathrm{c}$ is the limiting tensile resistance of the material, the crack can propagate with an angle $\theta_\mathrm{c}$, where $\theta_\mathrm{c}$ is fully determined by the stress state and, ahead of the crack tip, is always $-\uppi/2 < \theta_\mathrm{c} \le 0$ (see Figures~\ref{fig:rtrwa}c and~\ref{fig:rtrwa}d). Alternatively, the crack can propagate also along the contact interface, if $\bar{ \tilde{\sigma}}_{zz}(-\alpha) = \lambda$, i.e.\ if the tensile stress along the direction normal to the surface is equal to the interface strength $\lambda$. We now investigate when one or the other scenario prevails.

In the process zone ahead of the crack tip, it is always $\bar{ \tilde{\sigma}}_{zz}(\xi) < \lambda$ (cf. Eq.~\ref{eq:trwa_sigma}): this implies that whenever in this region the principal stress $\bar{ \tilde{\sigma}}_1$ is larger than or equal to the material strength $\tilde{\sigma}_\mathrm{c}$, the crack propagates within the bulk with an angle $\theta_\mathrm{c} > -\uppi/2$.

When this is not the case, the stress state at $\xi=-\alpha$ determines if the crack propagates along the interface or within the bulk. Here, if the principal stress $\bar{ \tilde{\sigma}}_1(-\alpha)$ is smaller than the material strength $\tilde{\sigma}_\mathrm{c}$, the crack propagates along the interface. If instead is $\bar{ \tilde{\sigma}}_1(-\alpha) > \tilde{\sigma}_\mathrm{c}$, the crack propagates within the bulk if

\begin{align}\label{eq:Lambda_1}
    \frac{\bar{ \tilde{\sigma}}_{1}(-\alpha)}{\tilde{\sigma}_c} > \frac{\bar{ \tilde{\sigma}}_{zz}(-\alpha)}{\lambda} \textrm{ .}
\end{align}

The effects of friction on the criterion of Eq.~\ref{eq:Lambda_1} is then of particular interest. Larger values of $\mu$ imply in fact larger values of $\bar{ \tilde{\sigma}}_1(\xi)$ (see Eq.~\ref{eq:trwa_sigma} and Figure~\ref{fig:rtrwa}a), thus enhancing crack propagation within the bulk. A similar trend is also expected for increasing values of $T/\mu P$, although the effects of a variation of such parameter appear negligible compared to those of $\mu$ (cf. Figure~\ref{fig:rtrwa}b).

Note that, if the crack propagates along the interface, neither the rolling particle nor the surface are worn. If the crack propagates within the bulk, detachment of material from the surface takes place, and the particle grows in size.

\subsection{Effect of cylinder radius}\label{ssec:rradius}
The value of the radius of the cylinder affects the value of the contact width, through the load equation (Eq.~\ref{eq:load}) and the non-dimensionalization of Equations~\ref{eq:nondim}. The positive solution for $\alpha$ of Equation~\ref{eq:load} is

\begin{align} \label{eq:alpha}
    \alpha = \frac{2}{\uppi} \left[ 2 \lambda \sqrt{m^2-1} + \sqrt{ \uppi \tilde{P} + 4 \lambda^2 \left(m^2-1\right)} \right]
\end{align}
which scales with $R$ as (cf. non-dimensionalizations~\ref{eq:nondim_l} to~\ref{eq:nondim_P}, and~\ref{eq:nondim_P0})

\begin{align} \label{eq:alphaR}
    \alpha \sim \sigma_0R^{1/3} + \sqrt{PR^{-1/3}+\sigma_0R^{2/3}} \quad \textrm{.}
\end{align}
The contact width $a$ thus scales as (cf. non-dimensionalizations~\ref{eq:nondim_a})

\begin{align} \label{eq:aR}
    a \sim R^{2/3} \left( \sigma_0R^{1/3} + \sqrt{PR^{-1/3}+\sigma_0R^{2/3}} \right)
\end{align}
and it always increases for increasing values of $R$ (holding the other parameters unchanged). Simple examples are the cases of zero applied load ($P=0$), for which $a \sim R$, and the Hertzian limit ($\sigma_0=0$), for which $a \sim R^{1/2}$.

For a given set of parameters, an increase in the contact width leads to lower values of the principal stress $\bar{ \sigma}_{1}(\xi)$: following from Equations~\ref{eq:nondim_s}, the true principal stress at the interface is in fact expressed as

\begin{align} \label{eq:aR}
  \bar{\sigma}_{1}(\xi) = s \cdot \bar{ \tilde{\sigma}}_{1}(\xi) = \left( \frac{w E^{*2}}{R} \right)^{1/3} \cdot \bar{ \tilde{\sigma}}_{1}(\xi)
\end{align}
and when $R$ increases, the true principal stress $\bar{\sigma}_{1}(\xi)$ generally decreases. Ahead of the crack tip, a reduction in the true principal stress $\bar{\sigma}_{1}(\xi)$ implies that there exists a crossover value $R_\times$ such that, for all $R \ge R_\times$, $\bar{\sigma}_{1}(\xi) < \sigma_\mathrm{c}$, and the crack propagation is governed by the stress state for $\xi = -\alpha$.

In this case, the value of the true stress $\bar{\sigma}_{zz}(-\alpha)$ is not affected by a change in $R$, as it is always given by the limiting adhesive stresses $\sigma_0$:

\begin{align} \label{eq:aR}
  \bar{\sigma}_{zz}(-\alpha) = s \cdot \lambda = s \cdot \frac{\sigma_0}{s} = \sigma_0 \textrm{ .}
\end{align}
We can now rewrite the criterion of Eq.~\ref{eq:Lambda_1} in terms of true stresses:
\begin{align}\label{eq:Lambda_criterion_true}
    \frac{\bar{ \sigma}_{1}(-\alpha)}{\sigma_c} > \frac{\bar{ \sigma}_{zz}(-\alpha)}{\sigma_0} \textrm{ ,}
\end{align}
and the only value that changes in Eq.~\ref{eq:Lambda_criterion_true} when $R$ increases is $\bar{\sigma}_{1}(-\alpha)$, which decreases, and it is thus harder to satisfy the criterion for crack propagation within the bulk: the probability of crack propagation along the interface thus increases.

To recap, for small values of $R$, crack propagation is more likely within the bulk, and for large values of $R$, crack propagation is more likely along the interface. If a particle then exhibits continuous growth during the rolling motion in the wear process, such growth is expected to decrease over time (as $R$ increases).

\subsection{Effect of Maugis parameter}\label{ssec:rmaugis}

The Maugis parameter is (cf. Equations~\ref{eq:nondim_lambda} and~\ref{eq:nondim_s})

\begin{align} \label{eq:alpha}
    \lambda = \sigma_0 \left( \frac{R}{w E^{*2}} \right)^{1/3}
\end{align}
and it affects the magnitude of the tensile stresses at the trailing edge: larger values of $\lambda$ imply larger values of $\bar{ \tilde{\sigma}}_1(\xi=-\alpha)$ and a higher chance of reaching the material resistance $\tilde{\sigma}_\mathrm{c}$ (cf. criterion of Eq.~\ref{eq:Lambda_1}). Large, soft bodies with strong adhesion ($\lambda \to \infty$, JKR limit) favour then crack propagation within the bulk (instead of along the interface) more than small, hard bodies with low adhesion ($\lambda \to 0$, DMT limit).

\subsection{Effect of load}\label{ssec:rload}

An increase in the applied load $P$ alone does not lead to an increase of $\bar{ \tilde{\sigma}}_{zz}(\xi)$ (the values being capped by $\lambda$ which depends on the material and the geometry, not on the loading conditions), but it does allow for larger tensile stresses $\bar{ \tilde{\sigma}}_{xx}^\mathrm{t}(\xi)$ due to the tangential component of the force transmitted at the interface (see Eq.~\ref{eq:carter_xx}). This is a consequence of the assumption of Coulomb friction, and it leads to larger principal stresses $\bar{ \tilde{\sigma}}_{1}(\xi \to -\alpha)$ and thus a higher likelihood of damage within the bulk. The trend of $\theta_\mathrm{c} (\xi / a )$ does not change.

\subsection{Effect of friction}\label{ssec:rfric}

\paragraph{Coefficient of friction}
For a given value of $T / \mu P$, the effect of the coefficient of friction $\mu$ is simply to scale the tractions $X(\xi)$ (cf. Equations~\ref{eq:trac_carter}) and thus larger values of $\mu$ leads to larger stresses $\bar{ \tilde{\sigma}}_1(\xi = -\alpha)$ (see Figures~\ref{fig:rtrwa} and~\ref{fig:rtrna}), extending the tensile zone ahead of the trailing edge and favouring crack propagation within the bulk (as the trend of $\theta_\mathrm{c} (\xi / a )$ is not affected by the value of $\mu$, see Figure~\ref{fig:rtrwa}c).

\paragraph{Transmitted forces ratio}
For a given value of $\mu$, the ratio $T / \mu P$ between the transmitted tangential force and the maximum frictional force changes the trend in $\theta_\mathrm{c} (\xi / a )$. Larger values of the ratio gives a larger tensile zone ahead of the trailing edge and a softer decrease in the values of $\theta_\mathrm{c}$ with $\xi/\alpha$ (see Figures~\ref{fig:rtrwa} and~\ref{fig:rtrna}). This favours crack propagation within the bulk as large values of $\lvert \theta_\mathrm{c} \rvert$ are encountered in the process zone.

In the limiting case of $\mu \to \infty$, it is $T / \mu P \to 0$, $\psi \to 0$, $\lim_{\xi \to -\chi^-} \theta_\mathrm{c} \to -\uppi/2$ and $\lim_{\xi \to -\chi^+} \theta_\mathrm{c} \to -\uppi/4$. The crack propagation angle ahead of the trailing edge is thus the smallest. In the most general case is then $-\uppi/2 < \theta_\mathrm{c} (\xi \to \alpha^+) \le \uppi/4$.

\paragraph{Adhesive friction}
Throughout the manuscript we assumed Coulomb friction. This is not strictly true when adhesion is strong, as is the case at small scales, where the frictional force is proportional to the true contact area instead~\citep{bowden2001friction,mo2009friction} and the linear proportionality with the normal load is lost~\citep{mo2009friction}. In a first approximation, when such contribution prevails, we can replace the tractions of Equation~\ref{eq:trna_X} with the distribution

\begin{align}
  \tilde{X}_\mathrm{adh}(\xi) =
  \begin{cases}
    \tilde{\tau}_0 \quad &\lvert \xi \rvert \le \chi \\
    0 \quad &\lvert \xi \rvert > \chi
  \end{cases}
\end{align}
where $\tilde{\tau}_0$ is simply the stress due to the applied tangential force if it is equally distributed over the interaction width $2\chi$. The latter coincides with the true contact area in our case, defined as the area over which the two bodies interact~\citep{mo2009friction}. Note that in this case the tangential distribution is applied over the whole interaction area, as it is assumed that tangential forces are sustained also where the two bodies are not in direct contact, as long as they interact with one another~\citep{mo2009friction}. Equation~\ref{eq:forcesnd_T} becomes then

\begin{align}
  \tilde{T} = \int_{-\chi}^{\chi} \tilde{X}_\mathrm{adh}(\xi) \textrm{d}\xi \quad \textrm{.}
\end{align}
and, to prevent sliding and slip, is limited by Equations~\ref{eq:muforcesnd} which now read

\begin{align}
  \lvert \tilde{T} \rvert &\le 2 \chi \tilde{\tau}_{c} \\
  \lvert \tilde{X}_\mathrm{adh}(\xi) \rvert &\le \tilde{\tau}_{c} \textrm{ , } \forall \xi \in \left[-\chi,\chi\right]
\end{align}
where $\tilde{\tau}_{c}$ is a non-dimensional effective shear strength of the contact interface.

The stress distribution $\bar{ \tilde{\sigma}}_{xx}^\mathrm{t}(\xi)$ for a constant tangential load is~\citep{johnson1987contact}

\begin{equation} \label{eq:sigma_adh}
\begin{aligned}
  & \bar{\tilde{\sigma}}_{zz}^\mathrm{t}(\xi) = 0 \\
  & \bar{\tilde{\sigma}}_{xx}^\mathrm{t}(\xi) = \frac{2 \tilde{\tau}_0}{\uppi} \ln \left( \frac{\xi-\chi}{\xi+\chi} \right) \\
  & \bar{\tilde{\sigma}}_{xz}^\mathrm{t}(\xi) =
    \begin{cases}
    -\tilde{\tau}_0 \quad &\lvert \xi \rvert \le \chi \\
    0 \quad &\lvert \xi \rvert > \chi
  \end{cases}
\end{aligned}
\end{equation}
which leads to a positive infinite value of $\bar{ \tilde{\sigma}}_{xx}^\mathrm{t}(\xi = -\chi)$ (the singularity is negative for $\xi=\chi$). We argue that such singularity may result in local plastification, but not crack propagation, as the crack tip is further ahead at $\xi=-\alpha$.

\begin{figure*}
  \centering
  \begin{subfigure}{.49\textwidth}
    \centering
    \includegraphics[width=\textwidth]{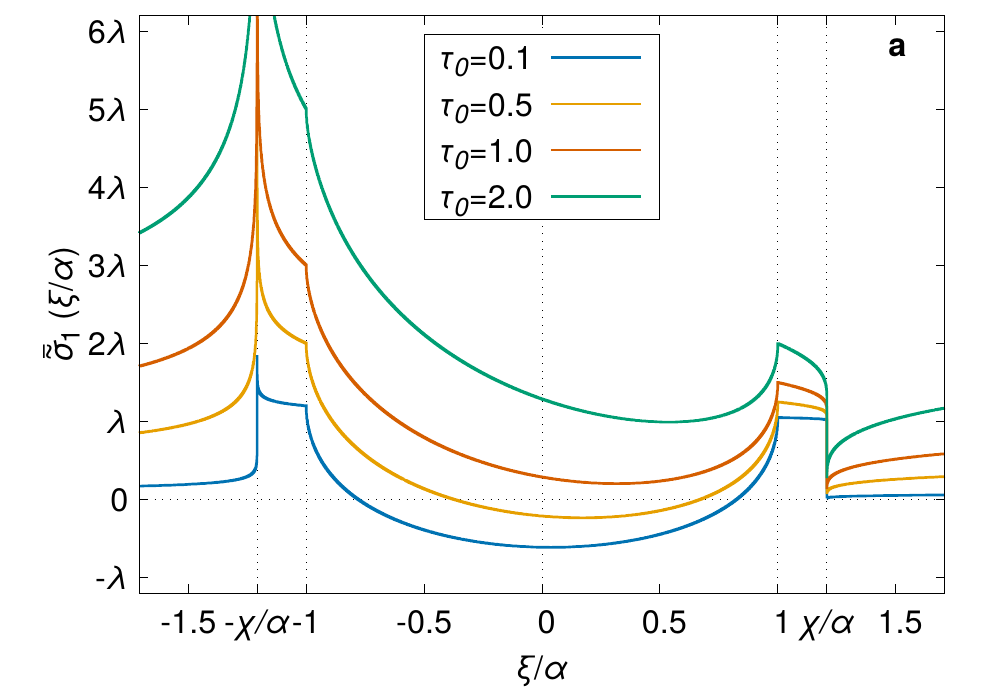}
  \end{subfigure}
  \begin{subfigure}{.49\textwidth}
    \centering
    \includegraphics[width=\textwidth]{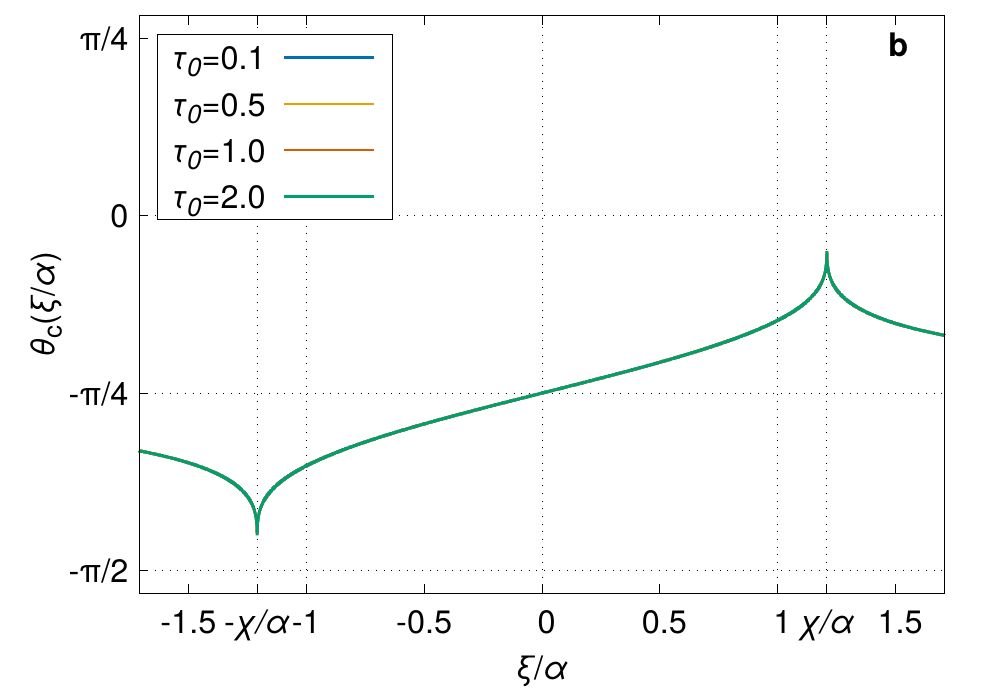}
  \end{subfigure}
  \caption{Maximum principal stress $\bar{ \tilde{\sigma}}_1(\xi)$ and crack propagation angle $\theta_\mathrm{c}$ along the interface in the case of tractive rolling with adhesion, with adhesive friction. The horizontal axis is scaled by $\alpha$. a) Effect of the applied constant tangential stress $\tilde{\tau}_0$ on $\bar{ \tilde{\sigma}}_1(\xi)$. b) The magnitude of the applied constant tangential stress $\tilde{\tau}_0$ has no effect on $\theta_\mathrm{c}$ (all curves superpose). In all the figures $\tilde{P}=\uppi / 4$ and $\lambda=1$.}
  \label{fig:adhfr}
\end{figure*}

The effects of the distribution of Equation~\ref{eq:sigma_adh} on the maximum principal stress and on the crack propagation angle are depicted in Figure~\ref{fig:adhfr}. With respect to the Coulomb friction case, $\bar{ \tilde{\sigma}}_1(\xi)$ changes significantly, and it shows a positive infinite value at $\xi = -\alpha$ independently of the value of $\tilde{\tau}_0$. The angle $\theta_\mathrm{c}$ is still in the interval $\left[ -\uppi/2, 0 \right]$, but its value is smaller at $\xi = \alpha$ than the Coulomb case. Adhesive friction is thus expected to lead to damage in a region closer to the surface.

Similarly to $\mu$ for the Coulomb friction case, $\tilde{\tau}_0$ has the effect of scaling the magnitude of $\bar{ \tilde{\sigma}}_1(\xi)$ leaving $\theta_\mathrm{c}$ unchanged.

In the most general case, the tangential tractions $\tilde{X}$ should include a term $\tilde{X}_{\mu}$ for the Coulomb friction and a second term $\tilde{X}_{\tilde{\tau}_0}$ for the adhesive friction, leading to an expression of the type \citep{carkner2010effect}

\begin{align}
  \tilde{X}(\xi) = \tilde{X}_{\mu} \left( \mu, \tilde{P}, \alpha \right) + \tilde{X}_{\tilde{\tau}_0} \left( \tilde{\tau}_0, \chi \right)
\end{align}
where one term or the other can prevail according to the material properties and the scale of the problem.

\subsection{Effect of pre-existing surface flaws}\label{ssec:reuler}
So far we assumed that the contact interface is continuous and homogeneous, the crack tip is located at $\xi = -\alpha$, and the crack propagates with an angle $\theta_\mathrm{c}$ if $\bar{ \tilde{\sigma}}_1(\xi) \ge \tilde{\sigma}_\mathrm{c}$ in the proximity of the crack tip. If it is always $\bar{ \tilde{\sigma}}_1(\xi) < \tilde{\sigma}_\mathrm{c}$, we can assume that rolling takes place without damaging the particle nor the surface, i.e.\ $\bar{ \tilde{\sigma}}_{zz}(\xi = -\alpha) = \lambda$, the interface adhesive resistance $\lambda$ is overcome, and the crack propagates along the contact interface. In other words, the tensile resistance $\lambda$ of the contact interface is reached before the tensile resistance $\tilde{\sigma}_\mathrm{c}$ of the bulk material.

On the other hand, the frame of reference moves with the contact interface: the surfaces of the cylinder and of the opposing surface flowing through it. This means that a point belonging to the surface goes through different stress states, from right to left in our stress diagrams, and the tensile resistance $\tilde{\sigma}_\mathrm{c}$ of the material can be reached at such points before they reach $\xi = -\alpha$. This has two consequences: First, if no surface flaw exists, the crack always propagates in the bulk with an angle $\theta_\mathrm{c} > -\uppi/2$, as soon as the material point reaches a high enough stress within the process zone. Second, if a surface flaw exists at the material point, a secondary crack may propagate before the point reaches $\xi = -\alpha$. The secondary crack still always propagates within the bulk (cf. $\theta_\mathrm{c}$ trends in previous sections and in \ref{sec:app1}), but at a lower angle, possibly leading to damage which is closer to the surface.

\subsection{Numerical validation}\label{sec:pf}

\paragraph{Methods}\label{ssec:methods}
Several numerical methods to model fracture are available in the literature, and are often divided into local and non-local approaches, depending on the crack propagation criterion. The extended finite element method~\citep{moes1999finite} and cohesive zone models~\citep{dugdale1960yielding,barenblatt1962mathematical} are examples of local models, where the criterion for crack propagation is applied at the tip of an existing crack or notch, and the crack is represented by a strong discontinuity in the material. In non-local methods, the crack is modelled instead as a continuous field: examples of such approaches are thick level sets~\citep{moes2011level} and phase-field approaches~\citep{francfort1998revisiting}. In the current work we adopted a phase-field approach, as it is based on a global energy minimization that takes into account of the energy needed to create new surfaces. This is possible because of the $\Gamma$-convergence of the method \citep{ambrosio1990approximation}, that allows to compute the fracture energy with an integral over the volume of the whole system instead of an integral over the crack surface (which is unknown). Another important feature of the approach is that the crack path is mesh independent.


The phase-field approach is based on~\cite{griffith1921vi} energy criterion for crack propagation, which states that, as the free energy of a system remains constant, the creation of new surfaces within a body takes place at the expenses of the potential energy of said body:

\begin{equation} \label{eq:pf_griffith}
\begin{aligned}
  \mathcal{E}(u_i,d) = \mathcal{E}_\mathrm{el}(u_i,d) + \mathcal{E}_\mathrm{frac}(d) - \mathcal{W}_\mathrm{ext}(u_i,d) \textrm{ ,}
\end{aligned}
\end{equation}
where $u_i$ and $d$ are the displacements and damage field, $\mathcal{E}$, $\mathcal{E}_\mathrm{el}$ and $\mathcal{E}_\mathrm{frac}$ indicate the free, elastic strain and fracture energies respectively, and $\mathcal{W}_\mathrm{ext}$ is the work done by the external forces on $u$. An increase in $\mathcal{E}_\mathrm{frac}$ is then possible only with concurrent reduction of the potential energy $\mathcal{E}_\mathrm{pot} = \mathcal{E}_\mathrm{el} - \mathcal{W}_\mathrm{ext}$. No kinetic energy appears in Eq.~\ref{eq:pf_griffith} as we are interested in a quasi-static approach, but extension of the method to include dynamics effects exist~\citep{li2016gradient,bleyer2017dynamic}. The introduction of a continuous damage field $d$ (with $d \in [0,1]$) in the phase-field approximation allows to represent the damage in the material, from intact ($d=0$) to fully damaged material ($d=1$). The fracture energy is then expressed as

\begin{equation} \label{eq:pf_frac_en}
\begin{aligned}
  \mathcal{E}_\mathrm{frac}(d) = \int_\Omega w \Gamma(d) \mathrm{d}\Omega \textrm{ ,}
\end{aligned}
\end{equation}
where $w$ is the fracture energy, $\Omega$ is the investigated body, and $\Gamma(d)$ the crack density functional introduced by the phase-field theory. Different expressions of $\Gamma(d)$ have been developed~\citep{borden2014higher,tanne2018crack} -- in our simulations we adopt the second order functional \texttt{AT2}~\citep{tanne2018crack}:

\begin{equation} \label{eq:pf_at2}
\begin{aligned}
  \Gamma(d) = \frac{1}{4 l_0} \left( d^2 + 4l_0^2 \left| \nabla d \right|^2 \right)  \textrm{ ,}
\end{aligned}
\end{equation}
where $l_0$ is a regularization length scale and $\nabla$ is the gradient operator.
To allow the crack to propagate only under tension (and not under compression), the elastic strain energy density in the phase-field approach is expressed as

\begin{equation} \label{eq:pf_strain_en}
\begin{aligned}
  \Psi_\mathrm{el}(\varepsilon_{ij},d) = \left(1-d \right)^2 \Psi_\mathrm{el}^+(\varepsilon_{ij}) + \Psi_\mathrm{el}^-(\varepsilon_{ij}) \textrm{ ,}
\end{aligned}
\end{equation}
where~\citep{miehe2010phase}

\begin{equation} \label{eq:pf_split}
\begin{aligned}
  \Psi_\mathrm{el}^\pm(\varepsilon_{ij}) &= \frac{\lambda'}{2} \langle \varepsilon_{ii} \rangle^2_\pm + G \varepsilon_{\pm ii}^2 \\
\end{aligned} \textrm{ ,}
\end{equation}
with $\lambda'$ and $G$ the Lamé first parameter and the shear modulus respectively, $\langle \bullet \rangle_\pm = \left( \bullet \pm \left| \bullet \right| \right)/2$. Note that other splits of $\Psi_\mathrm{el}(\varepsilon_{ij},d)$ are possible~\citep{li2016gradient,bleyer2017dynamic}.
By substituting Eq.~\ref{eq:pf_frac_en} and~\ref{eq:pf_strain_en} in Eq.~\ref{eq:pf_griffith}, the variational formulation of the phase-field approach is obtained. Numerically, the finite element method is used to discretize the system, and the free energy is minimized for $(u_i,d)$ with a staggered scheme~\citep{pham2017experimental,ambati2015review}. For each load increment, first the linear elastic problem is solved for the given values of the damage $d$, obtaining the updated values of the displacements $u_i$. These values of $u_i$ are then used to solve the phase-field problem and update the values of $d$, that are in turn used in the next solution of the linear elastic problem, and so on, until the values of $d$ and $u_i$ converge. The load is then incremented again. The phase-field and the finite element methods are available in the finite element library Akantu~\citep{richart2015implementation}.

\begin{figure*}
  \begin{subfigure}{\textwidth}
    \centering
    \includegraphics[width=\textwidth]{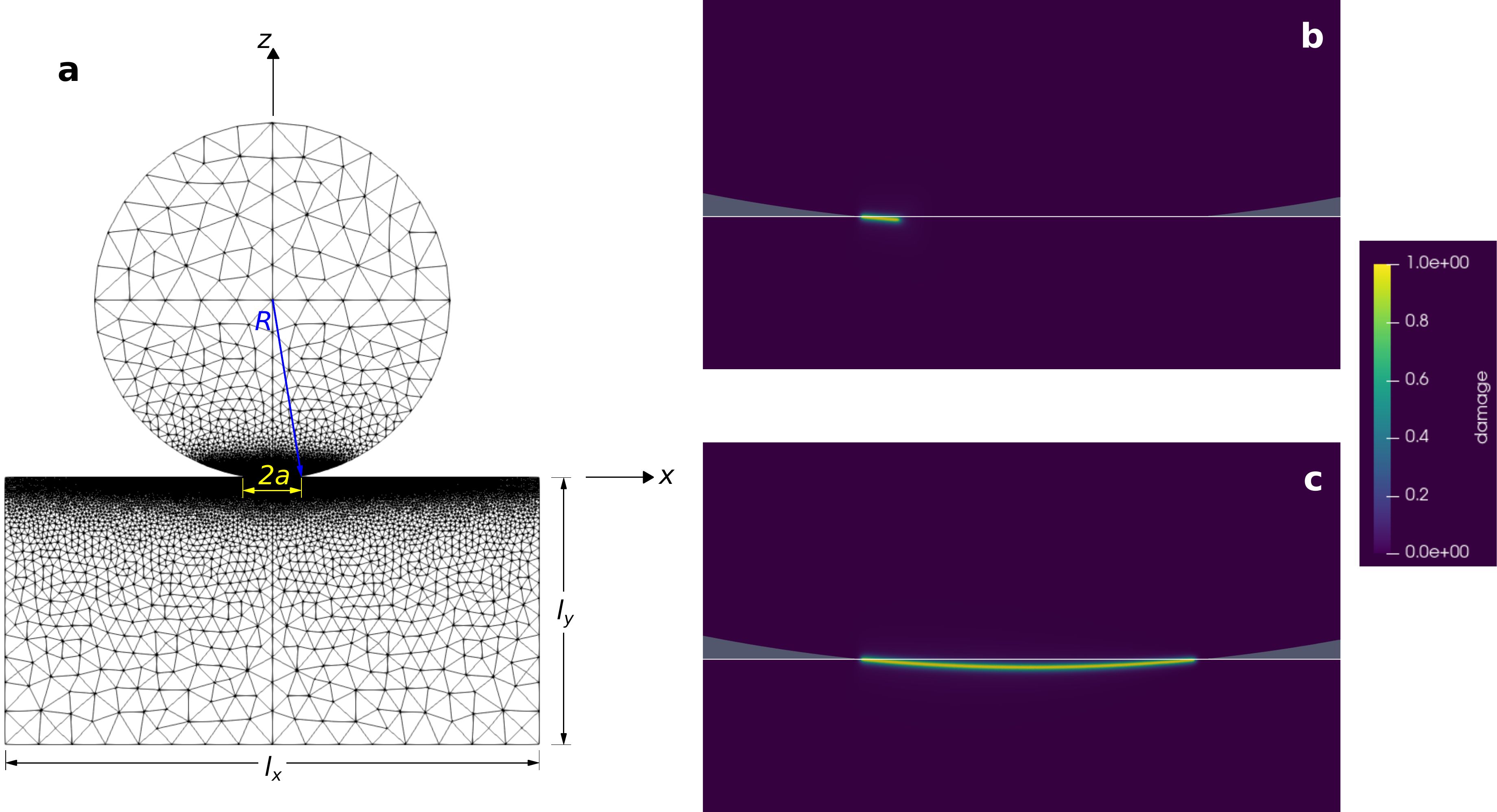}
  \end{subfigure}
  \caption{Discretized geometry (a) for the phase-field simulations and damage evolution (b-c). Note that the frame of reference of the finite element library Akantu follows the traditional continuum mechanics one, thus differing from the contact mechanics adopted throughout the manuscript (Figure~\ref{fig:fw}). b) Crack develops at the trailing edge of the contact interface and propagates with an angle $\theta_\mathrm{c} < 0$. c) At large loads, the crack bends back towards the leading edge. In both picture the white horizontal line refers to $z=0$.}
  \label{fig:pf_damage}
\end{figure*}

\paragraph{Simulation}\label{ssec:pf_sim}
The analytical prediction that the crack propagates within the bulk and not within the rolling particle has been tested with a numerical approach. A simplified geometry is then considered, where a cylinder in contact with a half-plane is modelled as one body, with no discontinuity along the contact interface (Figure~\ref{fig:pf_damage}a). The half-plane is represented by a large enough rectangular shape. The rolling motion is represented by applying a uniform horizontal displacements to the bottom and to the sides of the half-plane, and by fixing the center of mass of the cylinder. This choice allows the cylinder boundary to be stress-free. Details of the input parameters are provided in Table~\ref{tab:pf}. The fracture process is modelled by means of the phase-field approach described in the previous paragraph.

\begin{table*}
  \centering
  \begin{tabular}{c c}
    Parameter & Value \\[2pt]
    \hline \\[-8pt]
    $E$ & $1.0$ \\[2pt]
    $\nu$ & $0.25$ \\[2pt]
    $w$ & $0.1$ \\[2pt]
    $l_0$ & $0.01$ \\[2pt]
    $h_\mathrm{min}$ & $0.002$ \\[2pt]
    $R$ & $10$ \\[2pt]
    $l_x$ & $30$ \\[2pt]
    $l_y$ & $15$ \\[2pt]
    displacement rate & $0.01$ \\[2pt]
    number of steps & $2500$ \\[2pt]
    \hline \\[-8pt]
  \end{tabular}
  \caption{Parameters adopted for the numerical simulations. All values are in dimensionless units; lengths are non-dimensionalized by the half-width of contact. $h_\mathrm{min}$ is the value of the minimum mesh size, which extends over a stripe of thickness $1.8$ along the contact interface. This ensures the ratio $l_0 / h_\mathrm{min}=5$ which allows to correctly represent the crack thickness~\citep{tanne2018crack}. $l_x$ and $l_y$ are half-plane dimensions along the horizontal and vertical direction respectively.}
  \label{tab:pf}
\end{table*}

The results of the simulation are presented in Figures~\ref{fig:pf_damage}b-c. The crack propagates from the trailing to the leading edge, consistently with the rolling motion of a cylinder that advances towards positive values of the horizontal axis. The crack also propagates within the half-plane, as it is correctly predicted by our analytical model, until it reaches the leading edge. The phase-field model ensures that the crack path does not depend on the mesh.

\section{Conclusions}\label{sec:conclusions}
We have presented a framework to investigate wear formation and growth in a three-body contact configuration, where all bodies have the same material properties. This framework is inspired by the work of Maugis and sees the rolling motion as the opening of a crack at the trailing edge (and the closure of a crack at the leading edge). The direction of propagation of the opening crack determines from which body the material is removed in the wear process. We derive the full stress state at the surface in the free rolling and tractive rolling cases, with and without adhesion. If no tangential force is transmitted at the contact interface, the crack propagates along the interface and no wear is expected. If the resultant transmitted force has a tangential component, which opposes the direction of motion, the crack always propagates within the opposing surface, and never within the third-body. The effects of the particle size, the Maugis parameter, the frictional parameters, the adhesive friction, and surface flaws are also investigated. The trend of crack propagation within the bulk that we observe in our simplified model helps understanding recent observations revealing the growth of wear particles with time.

\begin{appendix}
  \setcounter{figure}{0}
  \renewcommand{\thesubsection}{A.\hspace{0.05em}\arabic{subsection}}

  \section{Further cases}\label{sec:app1}

  In this appendix, the cases of free rolling with and without adhesion and the case of tractive rolling without adhesion are discussed.

  \subsection{Calculation}


  When no adhesion is present ($\beta \to 0$ and $\lambda \to 0$)\footnote{ Note that both assumptions on $\beta$ and $\lambda$ are needed. If only vanishing adhesive stresses are assumed ($\lambda \to 0$), then $m \to \infty$ is admissible and the DMT limit is recovered. If only vanishing adhesive half-width is assumed ($\beta \to 0$), then $\lambda \to \infty$ is admissible and the JKR limit is recovered.}, the classical Hertzian case~\citep{hertz1882beruhrung} is recovered and Maugis normal tractions distribution reduces to  (cf. Equation~\ref{eq:maugis}, Figure~\ref{fig:trna_xx}):

  \begin{align}\label{eq:hertz}
    & \tilde{Z}(\xi) =
      \begin{cases}
        \frac{\tilde{p}_0}{\alpha} \sqrt{ \alpha^2-\xi^2} &\lvert \xi \rvert \le \alpha \\
        0 &\lvert \xi \rvert > \alpha
      \end{cases}
  \end{align}
  where $\tilde{p}_0$ is the pre-factor written in the classical Hertz convention (cf. Equation~\ref{eq:trwa_sigma}).

  \subsubsection{Free rolling without adhesion}\label{sssec:frna}
  If no tangential load is transmitted, i.e.\ $X(\xi)=0$, the rolling motion is free, the tractions at the interface are fully given by Equation~\ref{eq:hertz}, and the stresses $\bar{\tilde{\sigma}}_{ij}(\xi)$ are those due to pure Hertzian contact (cf. Equations~\ref{eq:sigma_red_cmp} and~\ref{eq:sxsz}):

  \begin{equation}\label{eq:frna_sigma}
    \begin{aligned}
      &\bar{ \tilde{\sigma}}_{zz}(\xi) = \bar{ \tilde{\sigma}}_{zz}^\mathrm{n}(\xi) =
      \begin{cases}
        -\frac{\tilde{p}_0}{\alpha} \sqrt{ \alpha^2-\xi^2} &\lvert \xi \rvert \le \alpha \\
        0 &\lvert \xi \rvert > \alpha
      \end{cases} \\
      &\bar{ \tilde{\sigma}}_{xx}(\xi) = \bar{ \tilde{\sigma}}_{xx}^\mathrm{n}(\xi) = \bar{ \tilde{\sigma}}_{zz}(\xi)\\
      &\bar{ \tilde{\sigma}}_{xz}(\xi) = \bar{ \tilde{\sigma}}_{xz}^\mathrm{n}(\xi) = 0
    \end{aligned} \quad \textrm{.}
  \end{equation}
  $\bar{ \tilde{\sigma}}_{xx}(\xi)$ and $\bar{ \tilde{\sigma}}_{zz}(\xi)$ are then also principal stresses (cf. Equation~\ref{eq:sigma_pr}), the stress state is hydrostatic, and all directions are principal directions.

  \subsubsection{Tractive rolling without adhesion}\label{sssec:trna}

\begin{figure*}
    \centering
    \begin{subfigure}{.49\textwidth}
      \centering
      \includegraphics[width=\textwidth]{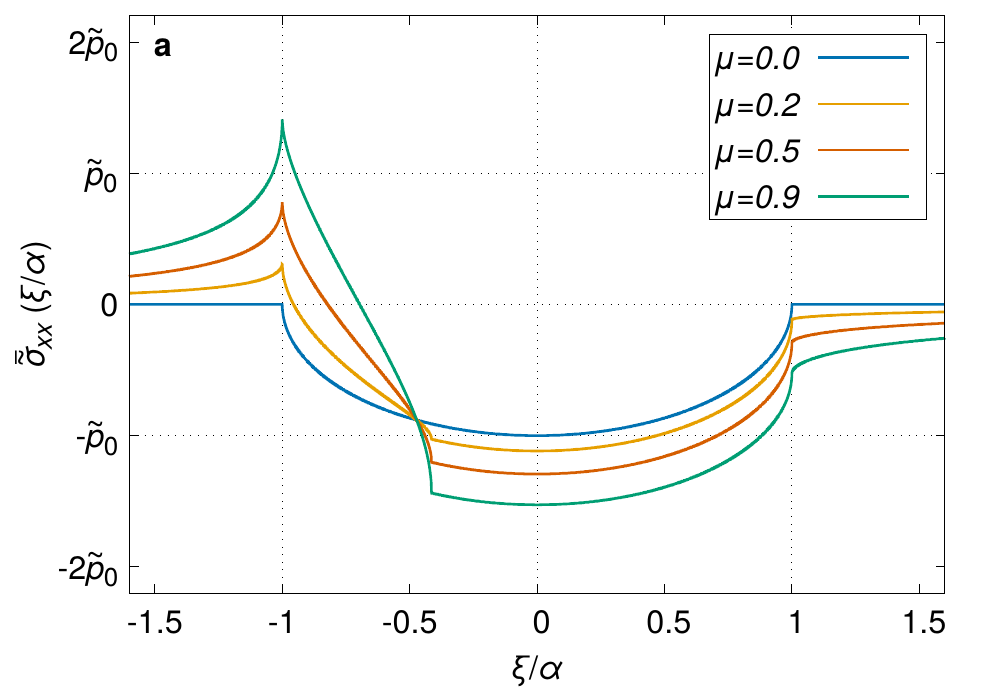}
    \end{subfigure}
    \begin{subfigure}{.49\textwidth}
      \centering
      \includegraphics[width=\textwidth]{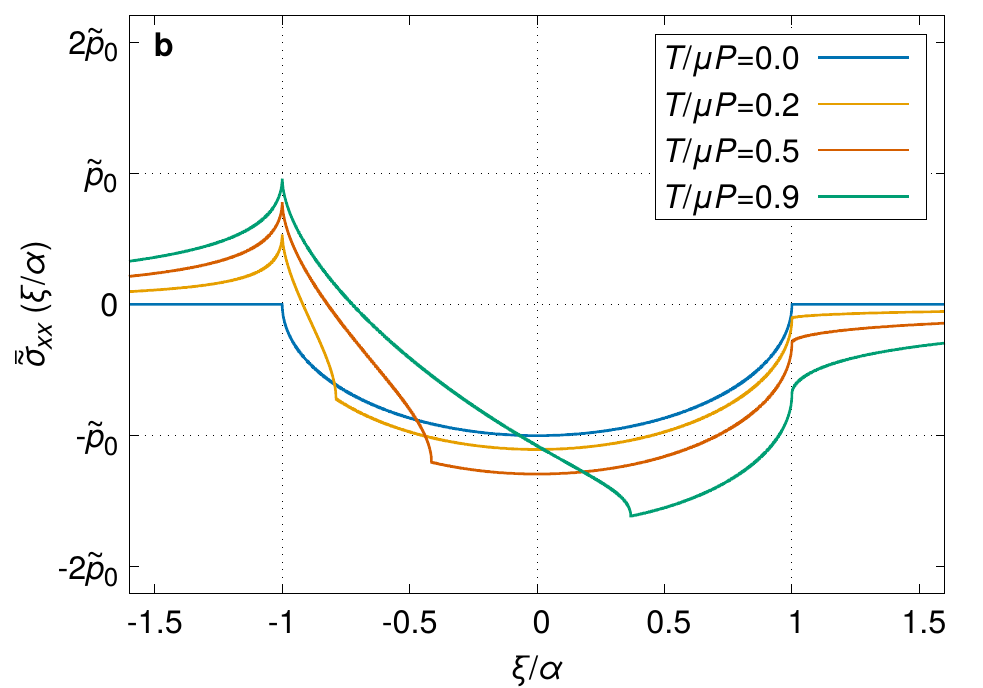}
    \end{subfigure}
    \caption{Stress $\bar{ \tilde{\sigma}}_{xx}(\xi)$ in the case of tractive rolling without adhesion. The horizontal axis is scaled by $\alpha$. a) Effect of the friction coefficient $\mu$ for constant values of the ratio between the transmitted tangential force and the limiting frictional force (here: $T / \mu P=0.5$). b) Effect of the ratio  $T / \mu P$ between the transmitted tangential force and the limiting frictional force for constant values of the coefficient of friction (here: $\mu=0.5$). In all the figures $\tilde{P}=\uppi / 4$ and thus $\alpha=1$ (cf. Equation~\ref{eq:load}, with $m \to 1$, $\lambda \to 0$ without adhesion.). $\mu=0$ coincides with the Hertzian case, for which also holds $\bar{ \tilde{\sigma}}_{zz}(\xi)=\bar{ \tilde{\sigma}}_{xx}(\xi)$.}
    \label{fig:trna_xx}
  \end{figure*}

  When the transmitted force has a tangential component, the stress state is given by Equations~\ref{eq:sigma_red_cmp}, with $\bar{\tilde{\sigma}}_{ij}^\mathrm{n}(\xi)$ given by Equations~\ref{eq:frna_sigma} in absence of adhesion and $\bar{\tilde{\sigma}}_{ij}^\mathrm{t}(\xi)$ by Equations~\ref{eq:carter} (see Figure~\ref{fig:trna_xx}):

  \begin{equation}\label{eq:trna_sigma}
    \begin{aligned}
      \bar{ \tilde{\sigma}}_{zz}(\xi) &=
      \begin{cases}
        -\frac{\tilde{p}_0}{\alpha} \sqrt{ \alpha^2-\xi^2} &\lvert \xi \rvert \le \alpha \\
        0 &\lvert \xi \rvert > \alpha
      \end{cases}\\
      \bar{ \tilde{\sigma}}_{xx}(\xi) &=
      \begin{cases}
        -\frac{\tilde{p}_0}{\alpha} \left( \sqrt{ \alpha^2 - \xi^2 } -2 \mu \xi \right) &\lvert \xi \rvert \le \alpha \\
        -2 \mu \frac{\tilde{p}_0}{\alpha} \left( \xi - \sgn\left( \xi \right) \sqrt{ \xi^2-\alpha^2 } \right) &\lvert \xi \rvert > \alpha
      \end{cases} +\\
      &+\begin{cases}
        2 \mu \frac{\tilde{p}_0}{\alpha} (\xi - \psi) &\lvert \xi-\psi \rvert \le \rho \\
        2 \mu \frac{\tilde{p}_0}{\alpha} \left[ \xi - \psi - \sgn\left( \xi - \psi \right) \sqrt{ \left(\xi -\psi\right)^2 - \rho^2 }\right] &\lvert \xi-\psi \rvert > \rho
      \end{cases}\\
      \bar{ \tilde{\sigma}}_{xz}(\xi) &=
      \begin{cases}
        -\mu \frac{\tilde{p}_0}{\alpha} \sqrt{ \alpha^2 - \xi^2 } &\lvert \xi \rvert \le \alpha \\
        0 &\lvert \xi \rvert > \alpha
      \end{cases} +\\
      &+\begin{cases}
        \mu \frac{\tilde{p}_0}{\alpha} \sqrt{ \rho^2 - \left(\xi-\psi\right)^2 } &\lvert \xi-\psi \rvert \le \rho \\
        0 &\lvert \xi - \psi \rvert > \rho
      \end{cases}
    \end{aligned} \textrm{ .}
  \end{equation}

  Note that in this case, besides the non-dimensionalized load $\tilde{P}$, (as for free rolling without adhesion, Section~\ref{sssec:frna}), it is necessary to know $\mu$ and $T/ \mu P$ to determine the stresses $\bar{ \tilde{\sigma}}_{ij}(\xi)$.

  \subsubsection{Free rolling with adhesion}\label{sssec:frwa}

  In the case of adhesion, the general expression of the Maugis tractions of Equations~\ref{eq:maugis} is used. As in Section~\ref{sssec:frna}, no tangential traction is transmitted and is $X(\xi)=0$. The surface stresses are then exactly those of Equations~\ref{eq:trna_eq} and are fully determined once the non-dimensionalized load $\tilde{P}$ and the Maugis parameter $\lambda$ are known. Also in this case $\bar{ \tilde{\sigma}}_{xx}(\xi)$ and $\bar{ \tilde{\sigma}}_{zz}(\xi)$ are principal stresses (cf. Equation~\ref{eq:sigma_pr}), the stress state is hydrostatic, and all directions are principal directions.

  \subsection{Results and discussion}

  \subsection{Free rolling without adhesion}\label{ssec:rfrna}

  From the stress state derived in section~\ref{sssec:frna}, at the trailing edge
  is $ \bar{ \tilde{\sigma}}_{xx}(-\alpha) = \bar{ \tilde{\sigma}}_{zz}(-\alpha) = \bar{ \tilde{\sigma}}_{xz}(-\alpha) = 0$. As all directions are principal, the crack will open along the weakest plane -- which normally is the contact interface.

  \subsection{Tractive rolling without adhesion}\label{ssec:rtrna}
\begin{figure*}
    \centering
    \begin{subfigure}{.49\textwidth}
      \centering
      \includegraphics[width=\textwidth]{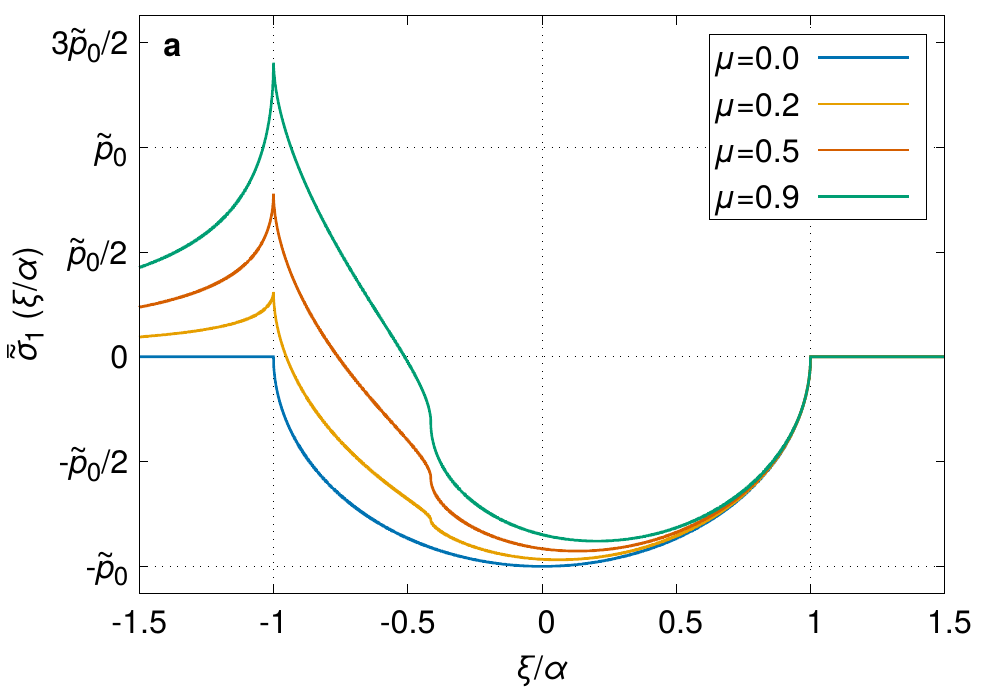}
    \end{subfigure}
    \begin{subfigure}{.49\textwidth}
      \centering
      \includegraphics[width=\textwidth]{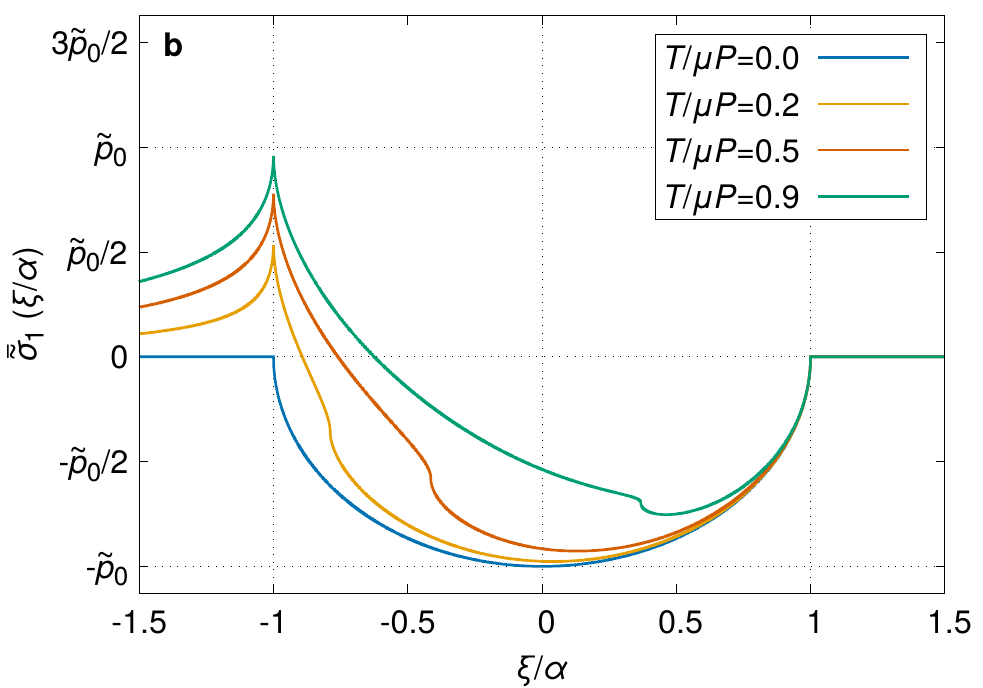}
    \end{subfigure}
    \begin{subfigure}{.49\textwidth}
      \centering
      \includegraphics[width=\textwidth]{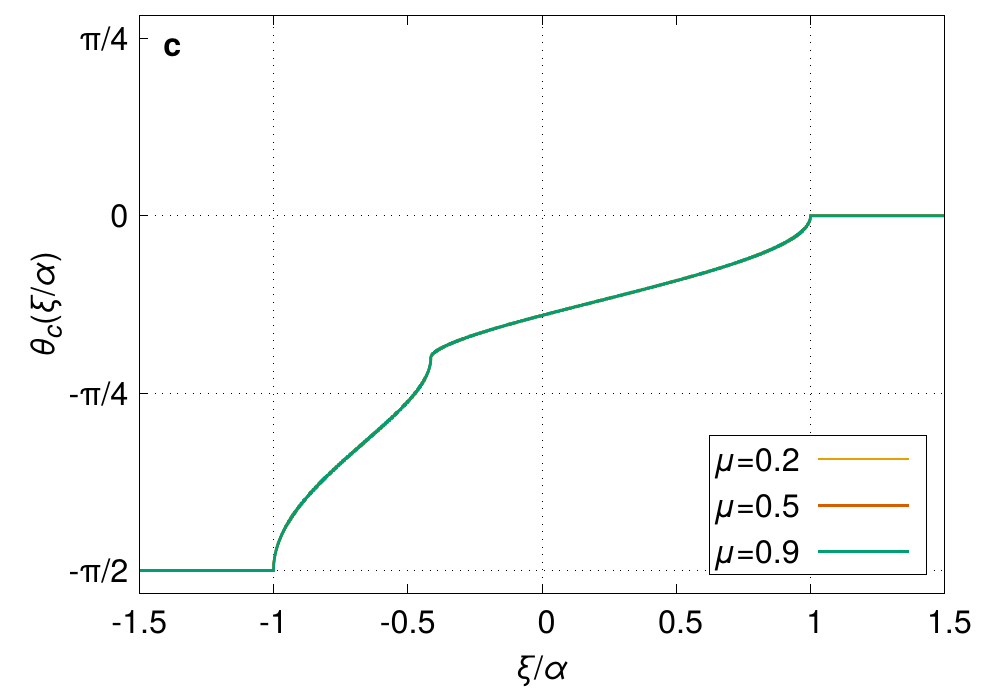}
    \end{subfigure}
    \begin{subfigure}{.49\textwidth}
      \centering
      \includegraphics[width=\textwidth]{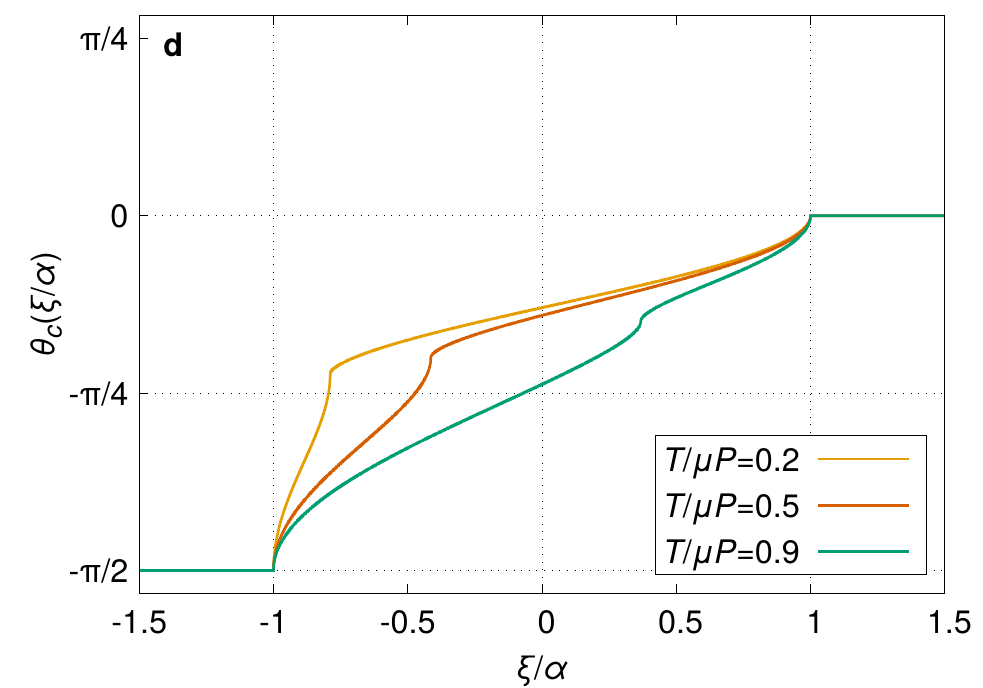}
    \end{subfigure}
    \caption{Maximum principal stress $\bar{ \tilde{\sigma}}_1(\xi)$ and crack propagation angle $\theta_\mathrm{c}$ along the interface in the case of tractive rolling without adhesion. The horizontal axis is scaled by $\alpha$. a) Effect of the friction coefficient $\mu$ on $\bar{ \tilde{\sigma}}_1(\xi)$ for constant values of the ratio between the transmitted tangential force and the limiting frictional force (here: $T / \mu P=0.5$). b) Effect of the ratio  $T / \mu P$  on $\bar{ \tilde{\sigma}}_1(\xi)$ for constant values of the coefficient of friction (here: $\mu=0.5$). c) The friction coefficient $\mu$ has no effect on $\theta_\mathrm{c}$ (all the curve superpose), for constant values of the ratio between the transmitted tangential force and the limiting frictional force (here: $T / \mu P=0.5$). d) Effect of the ratio  $T / \mu P$  on $\theta_\mathrm{c}$ for constant values of the coefficient of friction (here: $\mu=0.5$). In all the figures $\tilde{P}=\uppi / 4$ and thus $\alpha=1$ (cf. Equation~\ref{eq:load}).}
    \label{fig:rtrna}
  \end{figure*}

  Figures~\ref{fig:rtrna}a and~\ref{fig:rtrna}b show the maximum principal stress $\bar{ \tilde{\sigma}}_1(\xi)$ due to the stress fields derived in section~\ref{sssec:trna}. The stress is  maximum at the trailing edge ($\xi=-\alpha$), and is positive (tensile stress) over a region around the trailing edge. If $\bar{ \tilde{\sigma}}_1(-\alpha) > \tilde{\sigma}_\mathrm{c}$, $\tilde{\sigma}_\mathrm{c}$ being the maximum tensile stress that the material can sustain, the crack propagates with an angle $\theta_\mathrm{c}$. Such angle is fully determined by the stress state, and is represented in Figures~\ref{fig:rtrna}c and~\ref{fig:rtrna}d, and is always $-\uppi/2 < \theta_\mathrm{c} (\xi/a) \le 0$ ahead of the crack tip (i.e.\ for all $\xi > -\alpha$). The crack thus always propagates within the bottom material or along the contact interface, the actual angle depending on the size of the process zone.

  \subsection{Free rolling with adhesion}\label{ssec:rfrwa}
  From the stress state derived in section~\ref{sssec:frwa}, at the trailing edge is $ \bar{ \tilde{\sigma}}_{xx}(-\alpha) = \bar{ \tilde{\sigma}}_{zz}(-\alpha) = \lambda$ and $\bar{ \tilde{\sigma}}_{xz}(-\alpha) = 0$. Similarly to the case of free rolling without adhesion (Section~\ref{ssec:rfrna}), all directions are principal, the crack will open along the weakest plane -- which normally is the contact interface.

  \section{Maugis formalism}\label{asec:maugs}

  \renewcommand{\thesubsection}{B.\hspace{0.05em}\arabic{subsection}}

  Here we re-derive the equations presented in Sections~\ref{sec:theory} and~\ref{sec:calculation} following Maugis formalism. In such context, only the contact half-width $a$ is non-dimensionalized by the parameter $l$ (cf. Equations~\ref{eq:nondim_a} and~\ref{eq:nondim_l}), while the other lengths are non-dimensionalized by $a$ itself. Stresses are then plotted as a function of $x/a$. Note that such scaling for the stresses gives the same plots reported in the main body of the manuscript as $\xi/\alpha = x/a$. Also, the procedure in deriving the crack propagation angles $\theta_\mathrm{c}$ is not affected by the formalism, and all the arguments and the results reported in the main manuscript hold unchanged.

  In the following, we define $\tilde{x} \vcentcolon= x/a$, and all the parameters already introduced in the manuscript keep the same meaning.

  \subsection{Theory} \label{assec:theory}

  \subsubsection{Normal and tangential tractions} \label{assec:tractions}

  Equations~\ref{eq:forcesnd} and~\ref{eq:muforcesnd} in the Maugis formalism are:

  \begin{subequations} \label{eqA:forcesnd}
    \begin{align}
      & \tilde{P} = \int_{-m}^{m} \tilde{Z}(\tilde{x}) \textrm{d}\tilde{x} \\
      & \tilde{T} = \int_{-1}^{1} \tilde{X}(\tilde{x}) \textrm{d}\tilde{x}
    \end{align}
    \begin{align}
      & \lvert \tilde{T} \rvert \le \mu \tilde{P}  \label{eqA:muforces_g_nd}\\
      & \lvert \tilde{X}(\tilde{x}) \rvert \le \mu \tilde{Z}(\tilde{x})  \label{eqA:muforces_l_nd}
    \end{align}
  \end{subequations}

  Equations~\ref{eq:maugis} for the Maugis tractions become

  \begin{subequations} \label{eqA:maugis}
    \begin{align}
      & \tilde{Z}(\tilde{x}) =
        \begin{cases}
          \begin{aligned}
            & -\frac{ \alpha }{2} \sqrt{ 1-\tilde{x}^2} + \frac{2 \lambda}{\uppi} \arctan{ \sqrt{ \frac{m^2-1}{1-\tilde{x}^2} }  } &&\lvert\tilde{x} \rvert \le 1 \\
            & -\lambda && 1 < \lvert\tilde{x} \rvert \le m \\
            & 0 &&\lvert\tilde{x} \rvert > m
          \end{aligned}
        \end{cases}
    \end{align}
  \end{subequations}

  The distributions for the tangential load (Equations~\ref{eq:trac_carter}) are

  \begin{subequations} \label{eqA:trac_carter}
    \begin{align}
      & \tilde{X}'(\tilde{x}) =   \begin{cases}
        \begin{aligned}
          & \mu \frac{2\tilde{P}}{\uppi \alpha} \sqrt{ 1 -\tilde{x}^2 } &&\lvert\tilde{x} \rvert \le 1 \\
          & 0 &&\lvert\tilde{x} \rvert > 1
        \end{aligned}
      \end{cases} \label{eqA:trna_Xp}\\
      & \tilde{X}''(\tilde{x}) = \begin{cases}
        \begin{aligned}
          & - \mu \frac{2\tilde{P}}{\uppi \alpha} \sqrt{ \tilde{\rho}^2 - \left(\tilde{x}-\tilde{\psi}\right)^2 } &&\lvert\tilde{x}-\tilde{\psi} \rvert \le \tilde{\rho} \\
          & 0 &&\lvert\tilde{x} - \tilde{\psi} \rvert > \tilde{\rho}
        \end{aligned}
      \end{cases} \label{eqA:trna_Xpp}
    \end{align}
  \end{subequations}
  where $\tilde{\rho} = \frac{r}{a} = 1 - \tilde{\psi} = (1-\frac{T}{\mu P})^{1/2} \le 1$ is the non-dimensionalized half-width over which $\tilde{X}''(\tilde{x})$ is applied. The total tangential tractions at the interface are then~\citep{johnson1958effect}

  \begin{align} \label{eqA:trna_X}
    \tilde{X}(\tilde{x}) = \tilde{X}'(\tilde{x}) + \tilde{X}''(\tilde{x})
  \end{align}

  \subsubsection{Stress field}\label{asssec:stresses}

  Equations~\ref{eq:sigma_red} to~\ref{eq:sxsz} holds just by replacing $\xi$ with $\tilde{x}$.

  \paragraph{Stresses due to normal load}

  The surface stresses due to the normal component of the force transmitted at the interface are then

  \begin{subequations} \label{eqA:trna_eq}
    \begin{align}
      & \bar{ \tilde{\sigma}}_{xx}^\mathrm{n}(\tilde{x}) = \bar{ \tilde{\sigma}}_{zz}^\mathrm{n}(\tilde{x}) =
        \begin{cases}
          \begin{aligned}
            & \frac{ \alpha }{2} \sqrt{ 1-\tilde{x}^2} - \frac{2 \lambda}{\uppi} \arctan{ \sqrt{ \frac{m^2-1}{1-\tilde{x}^2} }  } &&\lvert\tilde{x} \rvert \le 1 \\
            & \lambda && 1 < \lvert\tilde{x} \rvert \le m \\
            & 0 &&\lvert\tilde{x} \rvert > m
          \end{aligned}
        \end{cases} \\
      & \bar{ \tilde{\sigma}}_{xz}^\mathrm{n}(\tilde{x}) = 0 \quad \textrm{.}
    \end{align}
  \end{subequations}

  To fully determine $\bar{ \tilde{\sigma}}_{ij}^\mathrm{n}(\tilde{x})$ is thus necessary and sufficient to know the non-dimensionalized normal component $\tilde{P}$ of the transmitted force and the Maugis parameter $\lambda$.

  \paragraph{Stresses due to tangential load}

  The surface stresses due to the tangential component of force transmitted at the interface are now (cf. Equations~\ref{eq:carter})

  \begin{subequations} \label{eqA:carter}
    \begin{align}
      & \bar{ \tilde{\sigma}}_{zz}^\mathrm{t}(\tilde{x}) = 0  \label{eqA:carter_xz}\\
      \begin{split} \label{eqA:carter_xz}
        & \bar{ \tilde{\sigma}}_{xz}^\mathrm{t}(\tilde{x}) = - \tilde{X}(\tilde{x}) = - \tilde{X}'(\tilde{x}) - \tilde{X}''(\tilde{x}) = \\
        &=\begin{cases}
          \begin{aligned}
            & -\mu \tilde{p}_0 \sqrt{ 1 -\tilde{x}^2 } &&\lvert\tilde{x} \rvert \le 1 \\
            & 0 &&\lvert\tilde{x} \rvert > 1
          \end{aligned}
        \end{cases} +
        \begin{cases}
          \begin{aligned}
            & \mu \tilde{p}_0 \sqrt{ \tilde{\rho}^2 - \left(\tilde{x}-\tilde{\psi}\right)^2 } &&\lvert\tilde{x}-\tilde{\psi} \rvert \le \tilde{\rho} \\
            & 0 &&\lvert\tilde{x} - \tilde{\psi} \rvert > \tilde{\rho}
          \end{aligned}
        \end{cases}
      \end{split}\\
      \begin{split}  \label{eqA:carter_xx}
        &\bar{ \tilde{\sigma}}_{xx}^\mathrm{t}(\tilde{x}) =  \bar{ \tilde{\sigma}}_{xx}^{\tilde{X}'}(\tilde{x}) + \bar{ \tilde{\sigma}}_{xx}^{\tilde{X}''}(\tilde{x}) = \\
        &=\begin{cases}
          \begin{aligned}
            & -2 \mu \tilde{p}_0\tilde{x} &&\lvert\tilde{x} \rvert \le 1 \\
            & -2 \mu \tilde{p}_0 \left(\tilde{x} - \sgn\left(\tilde{x} \right) \sqrt{\tilde{x}^2-1 } \right) &&\lvert\tilde{x} \rvert > 1
          \end{aligned}
        \end{cases} + \\
        &+\begin{cases}
          \begin{aligned}
            & 2 \mu \tilde{p}_0 (\tilde{x} - \tilde{\psi}) &&\lvert\tilde{x}-\tilde{\psi} \rvert \le \tilde{\rho} \\
            & 2 \mu \tilde{p}_0 \left[\tilde{x} - \tilde{\psi} - \sgn\left(\tilde{x} - \tilde{\psi} \right) \sqrt{ \left(\tilde{x} -\tilde{\psi}\right)^2 - \tilde{\rho}^2 }\right] &&\lvert\tilde{x}-\tilde{\psi} \rvert > \tilde{\rho}
          \end{aligned}
        \end{cases}
      \end{split}
    \end{align}
  \end{subequations}
  where the relation $\tilde{p}_0 = \frac{2 \tilde{P}}{ \uppi \alpha}$ has been used. For the derivation of Equations~\ref{eqA:carter_xx} we refer the reader to~\cite{johnson1987contact}.

  \subsection{Calculation}\label{assec:calculation}

  We report here the general cases of tractive rolling, with and without adhesion, as the cases of free rolling are straightforward once it is assumed $\mu=0$.

  \subsubsection{Tractive rolling without adhesion}\label{asssec:trna}

  Equations~\ref{eq:trna_sigma} in the Maugis formalism become

  \begin{equation}\label{eqA:trna_sigma}
    \begin{aligned}
      \bar{ \tilde{\sigma}}_{zz}(\tilde{x}) &=
      \begin{cases}
        -\tilde{p}_0 \sqrt{ 1 -\tilde{x}^2 } &\lvert\tilde{x} \rvert \le 1 \\
        0 &\lvert\tilde{x} \rvert > 1
      \end{cases}\\
      \bar{ \tilde{\sigma}}_{xx}(\tilde{x}) &=
      \begin{cases}
        -\tilde{p}_0 \left( \sqrt{ 1 -\tilde{x}^2 } -2 \mu\tilde{x} \right) &\lvert\tilde{x} \rvert \le 1 \\
        -2 \mu \tilde{p}_0 \left(\tilde{x} - \sgn\left(\tilde{x} \right) \sqrt{\tilde{x}^2-1 } \right) &\lvert\tilde{x} \rvert > 1
      \end{cases} +\\
      &+\begin{cases}
        2 \mu \tilde{p}_0 (\tilde{x} - \tilde{\psi}) &\lvert\tilde{x}-\tilde{\psi} \rvert \le \tilde{\rho} \\
        2 \mu \tilde{p}_0 \left[\tilde{x} - \tilde{\psi} - \sgn\left(\tilde{x} - \tilde{\psi} \right) \sqrt{ \left(\tilde{x} -\tilde{\psi}\right)^2 - \tilde{\rho}^2 }\right] &\lvert\tilde{x}-\tilde{\psi} \rvert > \tilde{\rho}
      \end{cases}\\
      \bar{ \tilde{\sigma}}_{xz}(\tilde{x}) &=
      \begin{cases}
        -\mu \tilde{p}_0 \sqrt{ 1 -\tilde{x}^2 } &\lvert\tilde{x} \rvert \le 1 \\
        0 &\lvert\tilde{x} \rvert > 1
      \end{cases} +\\
      &+\begin{cases}
        \mu \tilde{p}_0 \sqrt{ \tilde{\rho}^2 - \left(\tilde{x}-\tilde{\psi}\right)^2 } &\lvert\tilde{x}-\tilde{\psi} \rvert \le \tilde{\rho} \\
        0 &\lvert\tilde{x} - \tilde{\psi} \rvert > \tilde{\rho}
      \end{cases}
    \end{aligned}
  \end{equation}

  \subsubsection{Tractive rolling with adhesion}\label{asssec:trwa}

  Finally, Equations~\ref{eq:trwa_sigma} become

  \begin{equation}\label{eqA:trwa_sigma}
    \begin{aligned}
      \bar{ \tilde{\sigma}}_{zz}(\tilde{x}) &=
      \begin{cases}
        \tilde{p}_0 \sqrt{ 1-\tilde{x}^2} - \frac{2 \lambda}{\uppi} \arctan{ \sqrt{ \frac{m^2-1}{1-\tilde{x}^2} } } &\lvert\tilde{x} \rvert \le 1 \\
        \lambda & 1 < \lvert\tilde{x} \rvert \le m \\
        0 &\lvert\tilde{x} \rvert > m
      \end{cases} \\
      \bar{ \tilde{\sigma}}_{xx}(\tilde{x}) &=
      \begin{cases}
        -\tilde{p}_0 \left( \sqrt{ 1 -\tilde{x}^2 } -2 \mu\tilde{x} \right)- \frac{2 \lambda}{\uppi} \arctan{ \sqrt{ \frac{m^2-1}{1-\tilde{x}^2} } } &\lvert\tilde{x} \rvert \le 1 \\
        -2 \mu \tilde{p}_0 \left(\tilde{x} - \sgn\left(\tilde{x} \right) \sqrt{\tilde{x}^2-1 } \right) + \lambda &1 < \lvert\tilde{x} \rvert \le m  \\
        -2 \mu \tilde{p}_0 \left(\tilde{x} - \sgn\left(\tilde{x} \right) \sqrt{\tilde{x}^2-1 } \right) &\lvert\tilde{x} \rvert > m
      \end{cases} +\\
      &+\begin{cases}
        2 \mu \tilde{p}_0 (\tilde{x} - \tilde{\psi}) &\lvert\tilde{x}-\tilde{\psi} \rvert \le \tilde{\rho} \\
        2 \mu \tilde{p}_0 \left[\tilde{x} - \tilde{\psi} - \sgn\left(\tilde{x} - \tilde{\psi} \right) \sqrt{ \left(\tilde{x} -\tilde{\psi}\right)^2 - \tilde{\rho}^2 }\right] &\lvert\tilde{x}-\tilde{\psi} \rvert > \tilde{\rho}
      \end{cases}\\
      \bar{ \tilde{\sigma}}_{xz}(\tilde{x}) &=
      \begin{cases}
        -\mu \tilde{p}_0 \sqrt{ 1 -\tilde{x}^2 } &\lvert\tilde{x} \rvert \le 1 \\
        0 &\lvert\tilde{x} \rvert > 1
      \end{cases} +\\
      &+\begin{cases}
        \mu \tilde{p}_0 \sqrt{ \tilde{\rho}^2 - \left(\tilde{x}-\tilde{\psi}\right)^2 } &\lvert\tilde{x}-\tilde{\psi} \rvert \le \tilde{\rho} \\
        0 &\lvert\tilde{x} - \tilde{\psi} \rvert > \tilde{\rho}
      \end{cases}
    \end{aligned}
  \end{equation}
\end{appendix}
\section*{Acknowledgements}
E. M. thanks Mohit Pundir for the support in the usage of the phase-field implementation in Akantu \citep{richart2015implementation}.


\bibliography{mybibfile}

\begin{thebibliography}{61}
\expandafter\ifx\csname natexlab\endcsname\relax\def\natexlab#1{#1}\fi
\providecommand{\url}[1]{\texttt{#1}}
\providecommand{\href}[2]{#2}
\providecommand{\path}[1]{#1}
\providecommand{\DOIprefix}{doi:}
\providecommand{\ArXivprefix}{arXiv:}
\providecommand{\URLprefix}{URL: }
\providecommand{\Pubmedprefix}{pmid:}
\providecommand{\doi}[1]{\href{http://dx.doi.org/#1}{\path{#1}}}
\providecommand{\Pubmed}[1]{\href{pmid:#1}{\path{#1}}}
\providecommand{\bibinfo}[2]{#2}
\ifx\xfnm\relax \def\xfnm[#1]{\unskip,\space#1}\fi
\bibitem[{Aghababaei et~al.(2018)Aghababaei, Brink \&
  Molinari}]{aghababaei2018asperity}
\bibinfo{author}{Aghababaei, R.}, \bibinfo{author}{Brink, T.}, \&
  \bibinfo{author}{Molinari, J.-F.} (\bibinfo{year}{2018}).
\newblock \bibinfo{title}{Asperity-level origins of transition from mild to
  severe wear}.
\newblock {\it \bibinfo{journal}{Physical review letters}\/},  {\it
  \bibinfo{volume}{120}\/}, \bibinfo{pages}{186105}.
\bibitem[{Aghababaei et~al.(2016)Aghababaei, Warner \&
  Molinari}]{aghababaei2016critical}
\bibinfo{author}{Aghababaei, R.}, \bibinfo{author}{Warner, D.~H.}, \&
  \bibinfo{author}{Molinari, J.-F.} (\bibinfo{year}{2016}).
\newblock \bibinfo{title}{Critical length scale controls adhesive wear
  mechanisms}.
\newblock {\it \bibinfo{journal}{Nature Communications}\/},  {\it
  \bibinfo{volume}{7}\/}.
\bibitem[{Aghababaei et~al.(2017)Aghababaei, Warner \&
  Molinari}]{aghababaei2017debris}
\bibinfo{author}{Aghababaei, R.}, \bibinfo{author}{Warner, D.~H.}, \&
  \bibinfo{author}{Molinari, J.-F.} (\bibinfo{year}{2017}).
\newblock \bibinfo{title}{On the debris-level origins of adhesive wear}.
\newblock {\it \bibinfo{journal}{Proceedings of the National Academy of
  Sciences}\/},  {\it \bibinfo{volume}{114}\/}, \bibinfo{pages}{7935--7940}.
\bibitem[{Ambati et~al.(2015)Ambati, Gerasimov \&
  De~Lorenzis}]{ambati2015review}
\bibinfo{author}{Ambati, M.}, \bibinfo{author}{Gerasimov, T.}, \&
  \bibinfo{author}{De~Lorenzis, L.} (\bibinfo{year}{2015}).
\newblock \bibinfo{title}{A review on phase-field models of brittle fracture
  and a new fast hybrid formulation}.
\newblock {\it \bibinfo{journal}{Computational Mechanics}\/},  {\it
  \bibinfo{volume}{55}\/}, \bibinfo{pages}{383--405}.
\bibitem[{Ambrosio \& Tortorelli(1990)}]{ambrosio1990approximation}
\bibinfo{author}{Ambrosio, L.}, \& \bibinfo{author}{Tortorelli, V.~M.}
  (\bibinfo{year}{1990}).
\newblock \bibinfo{title}{Approximation of functional depending on jumps by
  elliptic functional via {$\Gamma$}-convergence}.
\newblock {\it \bibinfo{journal}{Communications on Pure and Applied
  Mathematics}\/},  {\it \bibinfo{volume}{43}\/}, \bibinfo{pages}{999--1036}.
\bibitem[{Archard(1953)}]{archard1953contact}
\bibinfo{author}{Archard, J.} (\bibinfo{year}{1953}).
\newblock \bibinfo{title}{Contact and rubbing of flat surfaces}.
\newblock {\it \bibinfo{journal}{Journal of Applied Physics}\/},  {\it
  \bibinfo{volume}{24}\/}, \bibinfo{pages}{981--988}.
\bibitem[{Baney \& Hui(1997)}]{baney1997cohesive}
\bibinfo{author}{Baney, J.}, \& \bibinfo{author}{Hui, C.-Y.}
  (\bibinfo{year}{1997}).
\newblock \bibinfo{title}{A cohesive zone model for the adhesion of cylinders}.
\newblock {\it \bibinfo{journal}{Journal of adhesion science and
  technology}\/},  {\it \bibinfo{volume}{11}\/}, \bibinfo{pages}{393--406}.
\bibitem[{Barenblatt et~al.(1962)}]{barenblatt1962mathematical}
\bibinfo{author}{Barenblatt, G.~I.} et~al. (\bibinfo{year}{1962}).
\newblock \bibinfo{title}{The mathematical theory of equilibrium cracks in
  brittle fracture}.
\newblock {\it \bibinfo{journal}{Advances in applied mechanics}\/},  {\it
  \bibinfo{volume}{7}\/}, \bibinfo{pages}{55--129}.
\bibitem[{Berthier et~al.(1988)Berthier, Vincent \&
  Godet}]{berthier1988velocity}
\bibinfo{author}{Berthier, Y.}, \bibinfo{author}{Vincent, L.}, \&
  \bibinfo{author}{Godet, M.} (\bibinfo{year}{1988}).
\newblock \bibinfo{title}{Velocity accommodation in fretting}.
\newblock {\it \bibinfo{journal}{Wear}\/},  {\it \bibinfo{volume}{125}\/},
  \bibinfo{pages}{25--38}.
\bibitem[{Bhaskaran et~al.(2010)Bhaskaran, Gotsmann, Sebastian, Drechsler,
  Lantz, Despont, Jaroenapibal, Carpick, Chen \&
  Sridharan}]{bhaskaran2010ultralow}
\bibinfo{author}{Bhaskaran, H.}, \bibinfo{author}{Gotsmann, B.},
  \bibinfo{author}{Sebastian, A.}, \bibinfo{author}{Drechsler, U.},
  \bibinfo{author}{Lantz, M.~A.}, \bibinfo{author}{Despont, M.},
  \bibinfo{author}{Jaroenapibal, P.}, \bibinfo{author}{Carpick, R.~W.},
  \bibinfo{author}{Chen, Y.}, \& \bibinfo{author}{Sridharan, K.}
  (\bibinfo{year}{2010}).
\newblock \bibinfo{title}{Ultralow nanoscale wear through atom-by-atom
  attrition in silicon-containing diamond-like carbon}.
\newblock {\it \bibinfo{journal}{Nature Nanotechnology}\/},  {\it
  \bibinfo{volume}{5}\/}, \bibinfo{pages}{181}.
\bibitem[{Bleyer et~al.(2017)Bleyer, Roux-Langlois \&
  Molinari}]{bleyer2017dynamic}
\bibinfo{author}{Bleyer, J.}, \bibinfo{author}{Roux-Langlois, C.}, \&
  \bibinfo{author}{Molinari, J.-F.} (\bibinfo{year}{2017}).
\newblock \bibinfo{title}{Dynamic crack propagation with a variational
  phase-field model: limiting speed, crack branching and velocity-toughening
  mechanisms}.
\newblock {\it \bibinfo{journal}{International Journal of Fracture}\/},  {\it
  \bibinfo{volume}{204}\/}, \bibinfo{pages}{79--100}.
\bibitem[{Boneh et~al.(2013)Boneh, Sagy \& Reches}]{boneh2013frictional}
\bibinfo{author}{Boneh, Y.}, \bibinfo{author}{Sagy, A.}, \&
  \bibinfo{author}{Reches, Z.} (\bibinfo{year}{2013}).
\newblock \bibinfo{title}{Frictional strength and wear-rate of carbonate faults
  during high-velocity, steady-state sliding}.
\newblock {\it \bibinfo{journal}{Earth and Planetary Science Letters}\/},  {\it
  \bibinfo{volume}{381}\/}, \bibinfo{pages}{127--137}.
\bibitem[{Borden et~al.(2014)Borden, Hughes, Landis \&
  Verhoosel}]{borden2014higher}
\bibinfo{author}{Borden, M.~J.}, \bibinfo{author}{Hughes, T.~J.},
  \bibinfo{author}{Landis, C.~M.}, \& \bibinfo{author}{Verhoosel, C.~V.}
  (\bibinfo{year}{2014}).
\newblock \bibinfo{title}{A higher-order phase-field model for brittle
  fracture: Formulation and analysis within the isogeometric analysis
  framework}.
\newblock {\it \bibinfo{journal}{Computer Methods in Applied Mechanics and
  Engineering}\/},  {\it \bibinfo{volume}{273}\/}, \bibinfo{pages}{100--118}.
\bibitem[{Bowden \& Tabor(2001)}]{bowden2001friction}
\bibinfo{author}{Bowden, F.~P.}, \& \bibinfo{author}{Tabor, D.}
  (\bibinfo{year}{2001}).
\newblock {\it \bibinfo{title}{The friction and lubrication of solids}\/}
  volume~\bibinfo{volume}{1}.
\newblock \bibinfo{publisher}{Oxford university press}.
\bibitem[{Brodsky et~al.(2011)Brodsky, Gilchrist, Sagy \&
  Collettini}]{brodsky2011faults}
\bibinfo{author}{Brodsky, E.~E.}, \bibinfo{author}{Gilchrist, J.~J.},
  \bibinfo{author}{Sagy, A.}, \& \bibinfo{author}{Collettini, C.}
  (\bibinfo{year}{2011}).
\newblock \bibinfo{title}{Faults smooth gradually as a function of slip}.
\newblock {\it \bibinfo{journal}{Earth and Planetary Science Letters}\/},  {\it
  \bibinfo{volume}{302}\/}, \bibinfo{pages}{185--193}.
\bibitem[{Carkner et~al.(2010)Carkner, Haw \& Mosey}]{carkner2010effect}
\bibinfo{author}{Carkner, C.~J.}, \bibinfo{author}{Haw, S.~M.}, \&
  \bibinfo{author}{Mosey, N.~J.} (\bibinfo{year}{2010}).
\newblock \bibinfo{title}{Effect of adhesive interactions on static friction at
  the atomic scale}.
\newblock {\it \bibinfo{journal}{Physical review letters}\/},  {\it
  \bibinfo{volume}{105}\/}, \bibinfo{pages}{056102}.
\bibitem[{Carter(1926)}]{carter1926action}
\bibinfo{author}{Carter, F.} (\bibinfo{year}{1926}).
\newblock \bibinfo{title}{On the action of a locomotive driving wheel}.
\newblock {\it \bibinfo{journal}{Proceedings of the Royal Society of London A:
  Mathematical, Physical and Engineering Sciences}\/},  {\it
  \bibinfo{volume}{112}\/}, \bibinfo{pages}{151--157}.
\bibitem[{Cattaneo(1938)}]{cattaneo1938sul}
\bibinfo{author}{Cattaneo, C.} (\bibinfo{year}{1938}).
\newblock \bibinfo{title}{Sul contatto di due corpi elastici: distribuzion
  locale degli sforzi}.
\newblock {\it \bibinfo{journal}{Rendiconti dell Accademia nazionale dei
  Lincei}\/},  {\it \bibinfo{volume}{27}\/}, \bibinfo{pages}{342--348,
  434--436, 474--478}.
\bibitem[{Chaudhury et~al.(1996)Chaudhury, Weaver, Hui \&
  Kramer}]{chaudhury1996adhesive}
\bibinfo{author}{Chaudhury, M.~K.}, \bibinfo{author}{Weaver, T.},
  \bibinfo{author}{Hui, C.}, \& \bibinfo{author}{Kramer, E.}
  (\bibinfo{year}{1996}).
\newblock \bibinfo{title}{Adhesive contact of cylindrical lens and a flat
  sheet}.
\newblock {\it \bibinfo{journal}{Journal of Applied Physics}\/},  {\it
  \bibinfo{volume}{80}\/}, \bibinfo{pages}{30--37}.
\bibitem[{Cocks(1962)}]{cocks1962interaction}
\bibinfo{author}{Cocks, M.} (\bibinfo{year}{1962}).
\newblock \bibinfo{title}{Interaction of sliding metal surfaces}.
\newblock {\it \bibinfo{journal}{Journal of Applied Physics}\/},  {\it
  \bibinfo{volume}{33}\/}, \bibinfo{pages}{2152--2161}.
\bibitem[{Da~Vinci(1478-1519)}]{davinci1519atlanticus}
\bibinfo{author}{Da~Vinci, L.} (\bibinfo{year}{1478-1519}).
\newblock {\it \bibinfo{title}{{C}odex {A}tlanticus}\/}.
\newblock \bibinfo{note}{URL: https://www.leonardodigitale.com/en/ Last visited
  on 24/02/2020.}
\bibitem[{Da~Vinci(1490-1499)}]{davincicodexI}
\bibinfo{author}{Da~Vinci, L.} (\bibinfo{year}{1490-1499}).
\newblock {\it \bibinfo{title}{{C}odex {M}adrid {I}}\/}.
\newblock \bibinfo{note}{URL: https://www.leonardodigitale.com/en/ Last visited
  on 24/02/2020.}
\bibitem[{Derjaguin et~al.(1975)Derjaguin, Muller \&
  Toporov}]{derjaguin1975effect}
\bibinfo{author}{Derjaguin, B.~V.}, \bibinfo{author}{Muller, V.~M.}, \&
  \bibinfo{author}{Toporov, Y.~P.} (\bibinfo{year}{1975}).
\newblock \bibinfo{title}{Effect of contact deformations on the adhesion of
  particles}.
\newblock {\it \bibinfo{journal}{Journal of Colloid and interface science}\/},
  {\it \bibinfo{volume}{53}\/}, \bibinfo{pages}{314--326}.
\bibitem[{Descartes \& Berthier(2002)}]{descartes2002rheology}
\bibinfo{author}{Descartes, S.}, \& \bibinfo{author}{Berthier, Y.}
  (\bibinfo{year}{2002}).
\newblock \bibinfo{title}{Rheology and flows of solid third bodies: background
  and application to an {MoS\textsubscript{1.6}} coating}.
\newblock {\it \bibinfo{journal}{Wear}\/},  {\it \bibinfo{volume}{252}\/},
  \bibinfo{pages}{546--556}.
\bibitem[{Dugdale(1960)}]{dugdale1960yielding}
\bibinfo{author}{Dugdale, D.~S.} (\bibinfo{year}{1960}).
\newblock \bibinfo{title}{Yielding of steel sheets containing slits}.
\newblock {\it \bibinfo{journal}{Journal of the Mechanics and Physics of
  Solids}\/},  {\it \bibinfo{volume}{8}\/}, \bibinfo{pages}{100--104}.
\bibitem[{Erdogan \& Sih(1963)}]{erdogan1963crack}
\bibinfo{author}{Erdogan, F.}, \& \bibinfo{author}{Sih, G.}
  (\bibinfo{year}{1963}).
\newblock \bibinfo{title}{On the crack extension in plates under plane loading
  and transverse shear}.
\newblock {\it \bibinfo{journal}{Journal of basic engineering}\/},  {\it
  \bibinfo{volume}{85}\/}, \bibinfo{pages}{519--525}.
\bibitem[{Fillot et~al.(2005)Fillot, Iordanoff \&
  Berthier}]{fillot2005simulation}
\bibinfo{author}{Fillot, N.}, \bibinfo{author}{Iordanoff, I.}, \&
  \bibinfo{author}{Berthier, Y.} (\bibinfo{year}{2005}).
\newblock \bibinfo{title}{Simulation of wear through mass balance in a dry
  contact}.
\newblock {\it \bibinfo{journal}{J. Trib.}\/},  {\it \bibinfo{volume}{127}\/},
  \bibinfo{pages}{230--237}.
\bibitem[{Fillot et~al.(2007{\natexlab{a}})Fillot, Iordanoff \&
  Berthier}]{fillot2007modelling}
\bibinfo{author}{Fillot, N.}, \bibinfo{author}{Iordanoff, I.}, \&
  \bibinfo{author}{Berthier, Y.} (\bibinfo{year}{2007}{\natexlab{a}}).
\newblock \bibinfo{title}{Modelling third body flows with a discrete element
  method—a tool for understanding wear with adhesive particles}.
\newblock {\it \bibinfo{journal}{Tribology International}\/},  {\it
  \bibinfo{volume}{40}\/}, \bibinfo{pages}{973--981}.
\bibitem[{Fillot et~al.(2007{\natexlab{b}})Fillot, Iordanoff \&
  Berthier}]{fillot2007wear}
\bibinfo{author}{Fillot, N.}, \bibinfo{author}{Iordanoff, I.}, \&
  \bibinfo{author}{Berthier, Y.} (\bibinfo{year}{2007}{\natexlab{b}}).
\newblock \bibinfo{title}{Wear modeling and the third body concept}.
\newblock {\it \bibinfo{journal}{Wear}\/},  {\it \bibinfo{volume}{262}\/},
  \bibinfo{pages}{949--957}.
\bibitem[{Francfort \& Marigo(1998)}]{francfort1998revisiting}
\bibinfo{author}{Francfort, G.~A.}, \& \bibinfo{author}{Marigo, J.-J.}
  (\bibinfo{year}{1998}).
\newblock \bibinfo{title}{Revisiting brittle fracture as an energy minimization
  problem}.
\newblock {\it \bibinfo{journal}{Journal of the Mechanics and Physics of
  Solids}\/},  {\it \bibinfo{volume}{46}\/}, \bibinfo{pages}{1319--1342}.
\bibitem[{Fr{\'{e}}rot et~al.(2018)Fr{\'{e}}rot, Aghababaei \&
  Molinari}]{Frerot2018}
\bibinfo{author}{Fr{\'{e}}rot, L.}, \bibinfo{author}{Aghababaei, R.}, \&
  \bibinfo{author}{Molinari, J.-F.} (\bibinfo{year}{2018}).
\newblock \bibinfo{title}{A mechanistic understanding of the wear coefficient:
  From single to multiple asperities contact}.
\newblock {\it \bibinfo{journal}{Journal of the Mechanics and Physics of
  Solids}\/},  {\it \bibinfo{volume}{114}\/}, \bibinfo{pages}{172--184}.
\bibitem[{Godet(1984)}]{godet1984third}
\bibinfo{author}{Godet, M.} (\bibinfo{year}{1984}).
\newblock \bibinfo{title}{The third-body approach: a mechanical view of wear}.
\newblock {\it \bibinfo{journal}{Wear}\/},  {\it \bibinfo{volume}{100}\/},
  \bibinfo{pages}{437--452}.
\bibitem[{Griffith(1921)}]{griffith1921vi}
\bibinfo{author}{Griffith, A.~A.} (\bibinfo{year}{1921}).
\newblock \bibinfo{title}{{VI.} {T}he phenomena of rupture and flow in solids}.
\newblock {\it \bibinfo{journal}{Philosophical transactions of the royal
  society of london. Series A, containing papers of a mathematical or physical
  character}\/},  {\it \bibinfo{volume}{221}\/}, \bibinfo{pages}{163--198}.
\bibitem[{Harris et~al.(2015)Harris, Curry, Pitenis, Rowe, Sidebottom, Sawyer
  \& Krick}]{harris2015wear}
\bibinfo{author}{Harris, K.~L.}, \bibinfo{author}{Curry, J.~F.},
  \bibinfo{author}{Pitenis, A.~A.}, \bibinfo{author}{Rowe, K.~G.},
  \bibinfo{author}{Sidebottom, M.~A.}, \bibinfo{author}{Sawyer, W.~G.}, \&
  \bibinfo{author}{Krick, B.~A.} (\bibinfo{year}{2015}).
\newblock \bibinfo{title}{Wear debris mobility, aligned surface roughness, and
  the low wear behavior of filled polytetrafluoroethylene}.
\newblock {\it \bibinfo{journal}{Tribology Letters}\/},  {\it
  \bibinfo{volume}{60}\/}, \bibinfo{pages}{2}.
\bibitem[{Hertz(1882)}]{hertz1882beruhrung}
\bibinfo{author}{Hertz, H.} (\bibinfo{year}{1882}).
\newblock \bibinfo{title}{{\"U}ber die {B}er{\"u}hrung fester elastischer
  {K}{\"o}rper}.
\newblock {\it \bibinfo{journal}{Journal f{\"u}r die reine und angewandte
  {M}athematik}\/},  {\it \bibinfo{volume}{92}\/}, \bibinfo{pages}{156--171}.
\bibitem[{Hintikka et~al.(2017)Hintikka, Lehtovaara \&
  M{\"a}ntyl{\"a}}]{hintikka2017third}
\bibinfo{author}{Hintikka, J.}, \bibinfo{author}{Lehtovaara, A.}, \&
  \bibinfo{author}{M{\"a}ntyl{\"a}, A.} (\bibinfo{year}{2017}).
\newblock \bibinfo{title}{Third particle ejection effects on wear with quenched
  and tempered steel fretting contact}.
\newblock {\it \bibinfo{journal}{Tribology Transactions}\/},  {\it
  \bibinfo{volume}{60}\/}, \bibinfo{pages}{70--78}.
\bibitem[{Holm(1946)}]{holm1946}
\bibinfo{author}{Holm, R.} (\bibinfo{year}{1946}).
\newblock {\it \bibinfo{title}{Electric contacts}\/}.
\newblock \bibinfo{publisher}{Almqvist and Wiksells, Stockholm}.
\bibitem[{Hutchings(2016)}]{hutchings2016leonardo}
\bibinfo{author}{Hutchings, I.~M.} (\bibinfo{year}{2016}).
\newblock \bibinfo{title}{{L}eonardo da {V}inci's studies of friction}.
\newblock {\it \bibinfo{journal}{Wear}\/},  {\it \bibinfo{volume}{360}\/},
  \bibinfo{pages}{51--66}.
\bibitem[{Jacobs \& Carpick(2013)}]{jacobs2013nanoscale}
\bibinfo{author}{Jacobs, T.~D.}, \& \bibinfo{author}{Carpick, R.~W.}
  (\bibinfo{year}{2013}).
\newblock \bibinfo{title}{Nanoscale wear as a stress-assisted chemical
  reaction}.
\newblock {\it \bibinfo{journal}{Nature Nanotechnology}\/},  {\it
  \bibinfo{volume}{8}\/}, \bibinfo{pages}{108}.
\bibitem[{Johnson(1958)}]{johnson1958effect}
\bibinfo{author}{Johnson, K.} (\bibinfo{year}{1958}).
\newblock \bibinfo{title}{The effect of a tangential contact force upon the
  rolling motion of an elastic sphere on a plane}.
\newblock {\it \bibinfo{journal}{Journal of Applied Mechanics}\/},  {\it
  \bibinfo{volume}{80}\/}, \bibinfo{pages}{339--346}.
\bibitem[{Johnson \& Greenwood(2008)}]{johnson2008maugis}
\bibinfo{author}{Johnson, K.}, \& \bibinfo{author}{Greenwood, J.}
  (\bibinfo{year}{2008}).
\newblock \bibinfo{title}{Maugis analysis of adhesive line contact}.
\newblock {\it \bibinfo{journal}{Journal of Physics D: Applied Physics}\/},
  {\it \bibinfo{volume}{41}\/}, \bibinfo{pages}{199802}.
\bibitem[{Johnson et~al.(1971)Johnson, Kendall \& Roberts}]{johnson1971surface}
\bibinfo{author}{Johnson, K.}, \bibinfo{author}{Kendall, K.}, \&
  \bibinfo{author}{Roberts, A.} (\bibinfo{year}{1971}).
\newblock \bibinfo{title}{Surface energy and the contact of elastic solids}.
\newblock {\it \bibinfo{journal}{Proceedings of the Royal Society of London A:
  Mathematical, Physical and Engineering Sciences}\/},  {\it
  \bibinfo{volume}{324}\/}, \bibinfo{pages}{301--313}.
\bibitem[{Johnson(1987)}]{johnson1987contact}
\bibinfo{author}{Johnson, K.~L.} (\bibinfo{year}{1987}).
\newblock {\it \bibinfo{title}{Contact mechanics}\/}.
\newblock \bibinfo{publisher}{Cambridge university press}.
\bibitem[{Li et~al.(2016)Li, Marigo, Guilbaud \& Potapov}]{li2016gradient}
\bibinfo{author}{Li, T.}, \bibinfo{author}{Marigo, J.-J.},
  \bibinfo{author}{Guilbaud, D.}, \& \bibinfo{author}{Potapov, S.}
  (\bibinfo{year}{2016}).
\newblock \bibinfo{title}{Gradient damage modeling of brittle fracture in an
  explicit dynamics context}.
\newblock {\it \bibinfo{journal}{International Journal for Numerical Methods in
  Engineering}\/},  {\it \bibinfo{volume}{108}\/}, \bibinfo{pages}{1381--1405}.
\bibitem[{Liu et~al.(2010{\natexlab{a}})Liu, Grierson, Moldovan, Notbohm, Li,
  Jaroenapibal, O'Connor, Sumant, Neelakantan, Carlisle
  et~al.}]{liu2010preventing}
\bibinfo{author}{Liu, J.}, \bibinfo{author}{Grierson, D.~S.},
  \bibinfo{author}{Moldovan, N.}, \bibinfo{author}{Notbohm, J.},
  \bibinfo{author}{Li, S.}, \bibinfo{author}{Jaroenapibal, P.},
  \bibinfo{author}{O'Connor, S.}, \bibinfo{author}{Sumant, A.},
  \bibinfo{author}{Neelakantan, N.}, \bibinfo{author}{Carlisle, J.~A.} et~al.
  (\bibinfo{year}{2010}{\natexlab{a}}).
\newblock \bibinfo{title}{Preventing nanoscale wear of atomic force microscopy
  tips through the use of monolithic ultrananocrystalline diamond probes}.
\newblock {\it \bibinfo{journal}{Small}\/},  {\it \bibinfo{volume}{6}\/},
  \bibinfo{pages}{1140--1149}.
\bibitem[{Liu et~al.(2010{\natexlab{b}})Liu, Notbohm, Carpick \&
  Turner}]{liu2010method}
\bibinfo{author}{Liu, J.}, \bibinfo{author}{Notbohm, J.~K.},
  \bibinfo{author}{Carpick, R.~W.}, \& \bibinfo{author}{Turner, K.~T.}
  (\bibinfo{year}{2010}{\natexlab{b}}).
\newblock \bibinfo{title}{Method for characterizing nanoscale wear of atomic
  force microscope tips}.
\newblock {\it \bibinfo{journal}{ACS Nano}\/},  {\it \bibinfo{volume}{4}\/},
  \bibinfo{pages}{3763--3772}.
\bibitem[{Maugis(1992)}]{maugis1992adhesion}
\bibinfo{author}{Maugis, D.} (\bibinfo{year}{1992}).
\newblock \bibinfo{title}{Adhesion of spheres: the {JKR}-{DMT} transition using
  a {D}ugdale model}.
\newblock {\it \bibinfo{journal}{Journal of Colloid and Interface Science}\/},
  {\it \bibinfo{volume}{150}\/}, \bibinfo{pages}{243--269}.
\bibitem[{Maugis(2000)}]{maugis2000modern}
\bibinfo{author}{Maugis, D.} (\bibinfo{year}{2000}).
\newblock \bibinfo{title}{Adhesion of solids: mechanical aspects}.
\newblock In \bibinfo{editor}{B.~Bhushan} (Ed.), {\it
  \bibinfo{booktitle}{Modern tribology handbook, two volume set}\/}
  chapter~\bibinfo{chapter}{4}. (pp. \bibinfo{pages}{163--204}).
\newblock \bibinfo{publisher}{CRC press}.
\bibitem[{Merkle \& Marks(2008)}]{merkle2008liquid}
\bibinfo{author}{Merkle, A.~P.}, \& \bibinfo{author}{Marks, L.~D.}
  (\bibinfo{year}{2008}).
\newblock \bibinfo{title}{Liquid-like tribology of gold studied by in situ
  {TEM}}.
\newblock {\it \bibinfo{journal}{Wear}\/},  {\it \bibinfo{volume}{265}\/},
  \bibinfo{pages}{1864--1869}.
\bibitem[{Miehe et~al.(2010)Miehe, Hofacker \& Welschinger}]{miehe2010phase}
\bibinfo{author}{Miehe, C.}, \bibinfo{author}{Hofacker, M.}, \&
  \bibinfo{author}{Welschinger, F.} (\bibinfo{year}{2010}).
\newblock \bibinfo{title}{A phase field model for rate-independent crack
  propagation: Robust algorithmic implementation based on operator splits}.
\newblock {\it \bibinfo{journal}{Computer Methods in Applied Mechanics and
  Engineering}\/},  {\it \bibinfo{volume}{199}\/}, \bibinfo{pages}{2765--2778}.
\bibitem[{Milanese et~al.(2019)Milanese, Brink, Aghababaei \&
  Molinari}]{milanese2019emergence}
\bibinfo{author}{Milanese, E.}, \bibinfo{author}{Brink, T.},
  \bibinfo{author}{Aghababaei, R.}, \& \bibinfo{author}{Molinari, J.-F.}
  (\bibinfo{year}{2019}).
\newblock \bibinfo{title}{Emergence of self-affine surfaces during adhesive
  wear}.
\newblock {\it \bibinfo{journal}{Nature communications}\/},  {\it
  \bibinfo{volume}{10}\/}, \bibinfo{pages}{1116}.
\bibitem[{Mo et~al.(2009)Mo, Turner \& Szlufarska}]{mo2009friction}
\bibinfo{author}{Mo, Y.}, \bibinfo{author}{Turner, K.~T.}, \&
  \bibinfo{author}{Szlufarska, I.} (\bibinfo{year}{2009}).
\newblock \bibinfo{title}{Friction laws at the nanoscale}.
\newblock {\it \bibinfo{journal}{Nature}\/},  {\it \bibinfo{volume}{457}\/},
  \bibinfo{pages}{1116}.
\bibitem[{Mo{\"e}s et~al.(1999)Mo{\"e}s, Dolbow \& Belytschko}]{moes1999finite}
\bibinfo{author}{Mo{\"e}s, N.}, \bibinfo{author}{Dolbow, J.}, \&
  \bibinfo{author}{Belytschko, T.} (\bibinfo{year}{1999}).
\newblock \bibinfo{title}{A finite element method for crack growth without
  remeshing}.
\newblock {\it \bibinfo{journal}{International journal for numerical methods in
  engineering}\/},  {\it \bibinfo{volume}{46}\/}, \bibinfo{pages}{131--150}.
\bibitem[{Mo{\"e}s et~al.(2011)Mo{\"e}s, Stolz, Bernard \&
  Chevaugeon}]{moes2011level}
\bibinfo{author}{Mo{\"e}s, N.}, \bibinfo{author}{Stolz, C.},
  \bibinfo{author}{Bernard, P.-E.}, \& \bibinfo{author}{Chevaugeon, N.}
  (\bibinfo{year}{2011}).
\newblock \bibinfo{title}{A level set based model for damage growth: the thick
  level set approach}.
\newblock {\it \bibinfo{journal}{International Journal for Numerical Methods in
  Engineering}\/},  {\it \bibinfo{volume}{86}\/}, \bibinfo{pages}{358--380}.
\bibitem[{Pham et~al.(2017)Pham, Ravi-Chandar \& Landis}]{pham2017experimental}
\bibinfo{author}{Pham, K.}, \bibinfo{author}{Ravi-Chandar, K.}, \&
  \bibinfo{author}{Landis, C.} (\bibinfo{year}{2017}).
\newblock \bibinfo{title}{Experimental validation of a phase-field model for
  fracture}.
\newblock {\it \bibinfo{journal}{International Journal of Fracture}\/},  {\it
  \bibinfo{volume}{205}\/}, \bibinfo{pages}{83--101}.
\bibitem[{Renouf et~al.(2011)Renouf, Massi, Fillot \&
  Saulot}]{renouf2011numerical}
\bibinfo{author}{Renouf, M.}, \bibinfo{author}{Massi, F.},
  \bibinfo{author}{Fillot, N.}, \& \bibinfo{author}{Saulot, A.}
  (\bibinfo{year}{2011}).
\newblock \bibinfo{title}{Numerical tribology of a dry contact}.
\newblock {\it \bibinfo{journal}{Tribology International}\/},  {\it
  \bibinfo{volume}{44}\/}, \bibinfo{pages}{834--844}.
\bibitem[{Richart \& Molinari(2015)}]{richart2015implementation}
\bibinfo{author}{Richart, N.}, \& \bibinfo{author}{Molinari, J.-F.}
  (\bibinfo{year}{2015}).
\newblock \bibinfo{title}{Implementation of a parallel finite-element library:
  test case on a non-local continuum damage model}.
\newblock {\it \bibinfo{journal}{Finite Elements in Analysis and Design}\/},
  {\it \bibinfo{volume}{100}\/}, \bibinfo{pages}{41--46}.
\bibitem[{Sorensen et~al.(1996)Sorensen, Jacobsen \&
  Stoltze}]{sorensen1996simulations}
\bibinfo{author}{Sorensen, M.}, \bibinfo{author}{Jacobsen, K.~W.}, \&
  \bibinfo{author}{Stoltze, P.} (\bibinfo{year}{1996}).
\newblock \bibinfo{title}{Simulations of atomic-scale sliding friction}.
\newblock {\it \bibinfo{journal}{Physical Review B}\/},  {\it
  \bibinfo{volume}{53}\/}, \bibinfo{pages}{2101--2113}.
\bibitem[{Tann{\'e} et~al.(2018)Tann{\'e}, Li, Bourdin, Marigo \&
  Maurini}]{tanne2018crack}
\bibinfo{author}{Tann{\'e}, E.}, \bibinfo{author}{Li, T.},
  \bibinfo{author}{Bourdin, B.}, \bibinfo{author}{Marigo, J.-J.}, \&
  \bibinfo{author}{Maurini, C.} (\bibinfo{year}{2018}).
\newblock \bibinfo{title}{Crack nucleation in variational phase-field models of
  brittle fracture}.
\newblock {\it \bibinfo{journal}{Journal of the Mechanics and Physics of
  Solids}\/},  {\it \bibinfo{volume}{110}\/}, \bibinfo{pages}{80--99}.
\bibitem[{Wu(2009)}]{wu2009adhesive}
\bibinfo{author}{Wu, J.-J.} (\bibinfo{year}{2009}).
\newblock \bibinfo{title}{Adhesive contact between a cylinder and a
  half-space}.
\newblock {\it \bibinfo{journal}{Journal of Physics D: Applied Physics}\/},
  {\it \bibinfo{volume}{42}\/}, \bibinfo{pages}{155302}.
\bibitem[{Yang et~al.(2016)Yang, Huang \& Shi}]{yang2016adhesion}
\bibinfo{author}{Yang, Y.}, \bibinfo{author}{Huang, L.}, \&
  \bibinfo{author}{Shi, Y.} (\bibinfo{year}{2016}).
\newblock \bibinfo{title}{Adhesion suppresses atomic wear in single-asperity
  sliding}.
\newblock {\it \bibinfo{journal}{Wear}\/},  {\it \bibinfo{volume}{352}\/},
  \bibinfo{pages}{31--41}.

\end{thebibliography}

\end{document}